\pdfoutput=1
\documentclass[12pt,a4paper]{article}
\usepackage{ifthen} % for conditional statements
\usepackage{tabularx}
\usepackage{booktabs}
\usepackage{multirow}
\usepackage{graphicx}
\usepackage{subcaption}
\usepackage{amsmath}
\usepackage{bm}
\usepackage{tikz}
\usepackage{setspace}
\usepackage{rotating}
\usepackage{float}

\newboolean{pdflatex}
\setboolean{pdflatex}{true} % False for eps figures 

\newboolean{articletitles}
\setboolean{articletitles}{true} % False removes titles in references

\newboolean{uprightparticles}
\setboolean{uprightparticles}{false} %True for upright particle symbols

\tikzset{boson/.style={decorate,decoration={coil,aspect=0}}}
\tikzset{vnode/.style={inner sep=0pt}}
\usetikzlibrary{decorations.markings, arrows.meta, decorations.pathmorphing}

\tikzset{
  midarrow/.style={
    decoration={
      markings,
      mark=at position 0.5 with {\arrow{Stealth}}
    },
    postaction={decorate}
  },
  doublearrow/.style={
        decoration={
            markings,
            mark=at position 0.25 with {\arrow{Stealth}},
            mark=at position 0.75 with {\arrow{Stealth}}
        },
        postaction={decorate}
    }
}

\def\paperauthors{LHCb collaboration} % Leave as is for PAPER, CONF and FIGURE
\def\paperasciititle{Study of the B0 to Lc Lcbar KS decay} % Set ASCII title here !! MAKE sure it's only ASCII characters !! 
\def\papertitle{Study of the \Bz \to \Lc \Lcbar \KS decay} % Latex formatted title
\def\paperkeywords{{High Energy Physics}, {LHCb}} % Comma separated list
\def\papercopyright{\the\year\ CERN for the benefit of the LHCb collaboration} % new since 9/Apr/2018
\def\paperlicence{CC BY 4.0 licence}
\def\paperlicenceurl{https://creativecommons.org/licenses/by/4.0/}

\newif\ifEnableSectionTOCLinks
\EnableSectionTOCLinksfalse % deactivated

\usepackage[top=1in, bottom=1.25in, left=1in, right=1in]{geometry}

\usepackage{microtype}
\usepackage{lineno}  % for line numbering during review
\usepackage{xspace} % To avoid problems with missing or double spaces after
\usepackage{caption} %these three command get the figure and table captions automatically small

\usepackage{graphicx}  % to include figures (can also use other packages)
\usepackage{color}
\usepackage{colortbl}
\graphicspath{{./figs/}} % Make Latex search fig subdir for figures
\usepackage{amsmath} % Adds a large collection of math symbols
\usepackage{amssymb}
\usepackage{amsfonts}
\usepackage{upgreek} % Adds in support for greek letters in roman typeset

\newcommand*\patchAmsMathEnvironmentForLineno[1]{%
\expandafter\let\csname old#1\expandafter\endcsname\csname #1\endcsname
\expandafter\let\csname oldend#1\expandafter\endcsname\csname
end#1\endcsname
 \renewenvironment{#1}%
   {\linenomath\csname old#1\endcsname}%
   {\csname oldend#1\endcsname\endlinenomath}%
}
\newcommand*\patchBothAmsMathEnvironmentsForLineno[1]{%
  \patchAmsMathEnvironmentForLineno{#1}%
  \patchAmsMathEnvironmentForLineno{#1*}%
}
\AtBeginDocument{%
\patchBothAmsMathEnvironmentsForLineno{equation}%
\patchBothAmsMathEnvironmentsForLineno{align}%
\patchBothAmsMathEnvironmentsForLineno{flalign}%
\patchBothAmsMathEnvironmentsForLineno{alignat}%
\patchBothAmsMathEnvironmentsForLineno{gather}%
\patchBothAmsMathEnvironmentsForLineno{multline}%
\patchBothAmsMathEnvironmentsForLineno{eqnarray}%
}

\usepackage[pdftex,
            pdfauthor={\paperauthors},
            pdftitle={\paperasciititle},
            pdfkeywords={\paperkeywords}]{hyperref}
\usepackage{hyperxmp}
\hypersetup{
    pdfcopyright={Copyright (C) \papercopyright},
    pdflicenseurl={\paperlicenceurl}
}
\usepackage[colorinlistoftodos,textsize=scriptsize]{todonotes}

\usepackage[bottom,flushmargin,hang,multiple]{footmisc}

\usepackage[all]{hypcap} % Internal hyperlinks to floats.

\usepackage{xspace} 
\usepackage{upgreek}

\def\lhcb   {\mbox{LHCb}\xspace}

\def\babar  {\mbox{BaBar}\xspace}

\def\besiii {\mbox{BESIII}\xspace}

\def\MagUp {\mbox{\em Mag\kern -0.05em Up}\xspace}

\ifthenelse{\boolean{uprightparticles}}%
{

 \def\Ppi         {\ensuremath{\uppi}\xspace}

 \def\PDelta      {\ensuremath{\Delta}\xspace}                 
 \def\PXi         {\ensuremath{\Xi}\xspace}                 
 \def\PLambda     {\ensuremath{\Lambda}\xspace}                 
 \def\PSigma      {\ensuremath{\Sigma}\xspace}                 
 \def\POmega      {\ensuremath{\Omega}\xspace}                 
 \def\PUpsilon    {\ensuremath{\Upsilon}\xspace}
 \let\oldPi\Pi
 \def\PPi         {\ensuremath{\oldPi}\xspace}

 \def\PB      {\ensuremath{\mathrm{B}}\xspace}                 
 \def\PD      {\ensuremath{\mathrm{D}}\xspace}                 

 \def\PK      {\ensuremath{\mathrm{K}}\xspace}                 
 \def\Pb      {\ensuremath{\mathrm{b}}\xspace}                 
 \def\Pc      {\ensuremath{\mathrm{c}}\xspace}                 
                  
 \def\Pe      {\ensuremath{\mathrm{e}}\xspace}                 
 \def\Pp      {\ensuremath{\mathrm{p}}\xspace}                 

 \def\Ps      {\ensuremath{\mathrm{s}}\xspace}

 \def\thebaroffset{0.0em}
}
{

 \def\Ppi         {\ensuremath{\pi}\xspace}

 \mathchardef\PDelta="7101
 \mathchardef\PXi="7104
 \mathchardef\PLambda="7103
 \mathchardef\PSigma="7106
 \mathchardef\POmega="710A
 \mathchardef\PUpsilon="7107
 \mathchardef\PPi="7105
 \def\PB      {\ensuremath{B}\xspace}                 
 \def\PD      {\ensuremath{D}\xspace}                 

 \def\PK      {\ensuremath{K}\xspace}                 
 \def\Pb      {\ensuremath{b}\xspace}                 
 \def\Pc      {\ensuremath{c}\xspace}                 
                  
 \def\Pe      {\ensuremath{e}\xspace}                 
 \def\Pp      {\ensuremath{p}\xspace}                 

 \def\Ps      {\ensuremath{s}\xspace}

 \def\thebaroffset{0.18em}
}
\newcommand{\offsetoverline}[2][\thebaroffset]{\kern #1\overline{\kern -#1 #2}}%

\makeatletter
\ifcase \@ptsize \relax% 10pt
  \newcommand{\miniscule}{\@setfontsize\miniscule{4}{5}}% \tiny: 5/6
\or% 11pt
  \newcommand{\miniscule}{\@setfontsize\miniscule{5}{6}}% \tiny: 6/7
\or% 12pt
  \newcommand{\miniscule}{\@setfontsize\miniscule{5}{6}}% \tiny: 6/7
\fi
\makeatother

\DeclareRobustCommand{\optbar}[1]{\shortstack{{\miniscule (\rule[.5ex]{1.25em}{.18mm})}
  \\ [-.7ex] $#1$}}

\def\epem       {{\ensuremath{\Pe^+\Pe^-}}\xspace}
\def\squark    {{\ensuremath{\Ps}}\xspace}

\def\cquark    {{\ensuremath{\Pc}}\xspace}
\def\cquarkbar {{\ensuremath{\overline \cquark}}\xspace}

\def\bquark    {{\ensuremath{\Pb}}\xspace}

\def\pion   {{\ensuremath{\Ppi}}\xspace}

\def\pip    {{\ensuremath{\pion^+}}\xspace}
\def\pim    {{\ensuremath{\pion^-}}\xspace}

\def\kaon    {{\ensuremath{\PK}}\xspace}
\def\KorKbar {\kern \thebaroffset\optbar{\kern -\thebaroffset \PK}{}\xspace}

\def\Kp      {{\ensuremath{\kaon^+}}\xspace}
\def\Km      {{\ensuremath{\kaon^-}}\xspace}

\def\KS      {{\ensuremath{\kaon^0_{\mathrm{S}}}}\xspace}

\def\D       {{\ensuremath{\PD}}\xspace}

\def\DorDbar {\kern \thebaroffset\optbar{\kern -\thebaroffset \PD}\xspace}

\def\Dp      {{\ensuremath{\D^+}}\xspace}
\def\Dm      {{\ensuremath{\D^-}}\xspace}

\def\DpDm    {\ensuremath{\Dp {\kern -0.16em \Dm}}\xspace}

\def\B       {{\ensuremath{\PB}}\xspace}

\def\BorBbar {\kern \thebaroffset\optbar{\kern -\thebaroffset \PB}\xspace}
\def\Bz      {{\ensuremath{\B^0}}\xspace}

\def\Bd      {{\ensuremath{\B^0}}\xspace}

\def\BdorBdbar {\kern \thebaroffset\optbar{\kern -\thebaroffset \Bd}\xspace}
\def\Bu      {{\ensuremath{\B^+}}\xspace}
\def\Bub     {{\ensuremath{\B^-}}\xspace}
\def\Bp      {{\ensuremath{\Bu}}\xspace}
\def\Bm      {{\ensuremath{\Bub}}\xspace}

\def\Bs      {{\ensuremath{\B^0_\squark}}\xspace}

\def\BsorBsbar {\kern \thebaroffset\optbar{\kern -\thebaroffset \Bs}\xspace}

\def\Y#1S{\ensuremath{\PUpsilon{(#1S)}}\xspace}

\def\proton      {{\ensuremath{\Pp}}\xspace}
\def\antiproton  {{\ensuremath{\overline \proton}}\xspace}

\def\Lz          {{\ensuremath{\PLambda}}\xspace}
\def\Lbar        {{\ensuremath{\offsetoverline{\PLambda}}}\xspace}
\def\LorLbar     {\kern \thebaroffset\optbar{\kern -\thebaroffset \PLambda}\xspace}

\def\Xires       {{\ensuremath{\PXi}}\xspace}

\def\myLc          {{\ensuremath{\Lz_\cquark}}\xspace}
\def\Lc          {{\ensuremath{\Lz^+_\cquark}}\xspace}
\def\Lcbar       {{\ensuremath{\Lbar{}^-_\cquark}}\xspace}

\def\Xic         {{\ensuremath{\Xires_\cquark}}\xspace}

\def\Xicp        {{\ensuremath{\Xires^+_\cquark}}\xspace}

\def\BF         {{\ensuremath{\mathcal{B}}}\xspace}

\newcommand{\decay}[2]{\ensuremath{\mathinner{#1\!\to #2}}\xspace}

\def\to                 {\ensuremath{\rightarrow}\xspace}

\def\eps   {{\ensuremath{\varepsilon}}\xspace}

\def\AT#1     {\ensuremath{A_{\mathrm{T}}^{#1}}\xspace}           % 2

\def\C#1      {\ensuremath{\mathcal{C}_{#1}}\xspace}                       % 9
\def\Cp#1     {\ensuremath{\mathcal{C}_{#1}^{'}}\xspace}                    % 7
\def\Ceff#1   {\ensuremath{\mathcal{C}_{#1}^{\mathrm{(eff)}}}\xspace}        % 9  
\def\Cpeff#1  {\ensuremath{\mathcal{C}_{#1}^{'\mathrm{(eff)}}}\xspace}       % 7
\def\Ope#1    {\ensuremath{\mathcal{O}_{#1}}\xspace}                       % 2
\def\Opep#1   {\ensuremath{\mathcal{O}_{#1}^{'}}\xspace}                    % 7

\newcommand{\nospaceunit}[1]{\ensuremath{\text{#1}}}       
\newcommand{\aunit}[1]{\ensuremath{\text{\,#1}}}       
\newcommand{\tev}{\aunit{Te\kern -0.1em V}\xspace}
\newcommand{\gev}{\aunit{Ge\kern -0.1em V}\xspace}
\newcommand{\mev}{\aunit{Me\kern -0.1em V}\xspace}
\newcommand{\kev}{\aunit{ke\kern -0.1em V}\xspace}
\newcommand{\ev}{\aunit{e\kern -0.1em V}\xspace}
 
\newcommand{\mevc}{\ensuremath{\aunit{Me\kern -0.1em V\!/}c}\xspace}
\newcommand{\gevc}{\ensuremath{\aunit{Ge\kern -0.1em V\!/}c}\xspace}
\newcommand{\mevcc}{\ensuremath{\aunit{Me\kern -0.1em V\!/}c^2}\xspace}
\newcommand{\gevcc}{\ensuremath{\aunit{Ge\kern -0.1em V\!/}c^2}\xspace}
\def\mum  {\ensuremath{\,\upmu\nospaceunit{m}}\xspace}

\def\fm   {\aunit{fm}\xspace}

\def\fb   {\ensuremath{\aunit{fb}}\xspace}
\def\invfb   {\ensuremath{\fb^{-1}}\xspace}

\def\ps   {\ensuremath{\aunit{ps}}\xspace}

\def\gsim{{~\raise.15em\hbox{$>$}\kern-.85em
          \lower.35em\hbox{$\sim$}~}\xspace}
\def\lsim{{~\raise.15em\hbox{$<$}\kern-.85em
          \lower.35em\hbox{$\sim$}~}\xspace}

\def\pt         {\ensuremath{p_{\mathrm{T}}}\xspace}

\def\ptot       {\ensuremath{p}\xspace}

\def\evtgen     {\mbox{\textsc{EvtGen}}\xspace}

\def\geant      {\mbox{\textsc{Geant4}}\xspace}

\def\photos     {\mbox{\textsc{Photos}}\xspace}

\def\pythia     {\mbox{\textsc{Pythia}}\xspace}

\def\tell1  {TELL1\xspace}
\def\ukl1   {UKL1\xspace}

\newcommand{\lhcborcid}[1]{\href{https://orcid.org/#1}{\hspace*{0.1em}\raisebox{-0.45ex}{\includegraphics[width=1em]{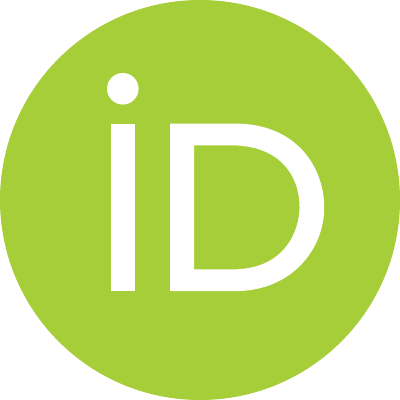}}}}

\hypersetup{
  colorlinks   = true, %Colours links instead of ugly boxes
  urlcolor     = blue, %Colour for external hyperlinks
  linkcolor    = blue, %Colour of internal links
  citecolor    = red   %Colour of citations
}

\ifEnableSectionTOCLinks
    \usepackage[explicit]{titlesec} % to change headings
    
    \let\oldcontentsline\contentsline
    \renewcommand

    \titleformat{\section}{\normalfont\Large\bf}{\hyperlink{tocsection.\thesection}{{\thesection} \parbox[t]{\dimexpr\textwidth-1pc}{#1}}}{1pc}{}

    \titleformat{\subsection}{\normalfont\bf}{\hyperlink{tocsubsection.\thesubsection}{{\thesubsection} \parbox[t]{\dimexpr\textwidth-1pc}{#1}}}{1pc}{}

    \titleformat{name=\section,numberless}[display]{}{}{0pt}{\normalfont\Huge\bfseries #1}
\fi

\usepackage{cite} % Allows for ranges in citations
\usepackage{LHCb/mciteplus}
\usepackage{longtable} % only for template; not usually to be used in PAPERs

\begin{document}

%%%%%%%%%%%%%%%%%%%%%%%%%
%%%%% Title     %%%%%%%%%
%%%%%%%%%%%%%%%%%%%%%%%%%
\renewcommand{\thefootnote}{\fnsymbol{footnote}}
\setcounter{footnote}{1}

% %%%%%%% CHOOSE TITLE PAGE--------
%\onecolumn
% \input{title-LHCb-PAPER}
% ===============================================================================
% Purpose: LHCb-PAPER journal paper title page template
% Author: 
% Created on: 2010-09-25
% ===============================================================================

%%%%%%%%%%%%%%%%%%%%%%%%%
%%%%%  TITLE PAGE  %%%%%%
%%%%%%%%%%%%%%%%%%%%%%%%%
\begin{titlepage}
\pagenumbering{roman}

% Header ---------------------------------------------------
\vspace*{-1.5cm}
\centerline{\large EUROPEAN ORGANIZATION FOR NUCLEAR RESEARCH (CERN)}
\vspace*{1.5cm}
\noindent
\begin{tabular*}{\linewidth}{lc@{\extracolsep{\fill}}r@{\extracolsep{0pt}}}
\ifthenelse{\boolean{pdflatex}}% Logo format choice
{\vspace*{-1.5cm}\mbox{\!\!\!\includegraphics[width=.14\textwidth]{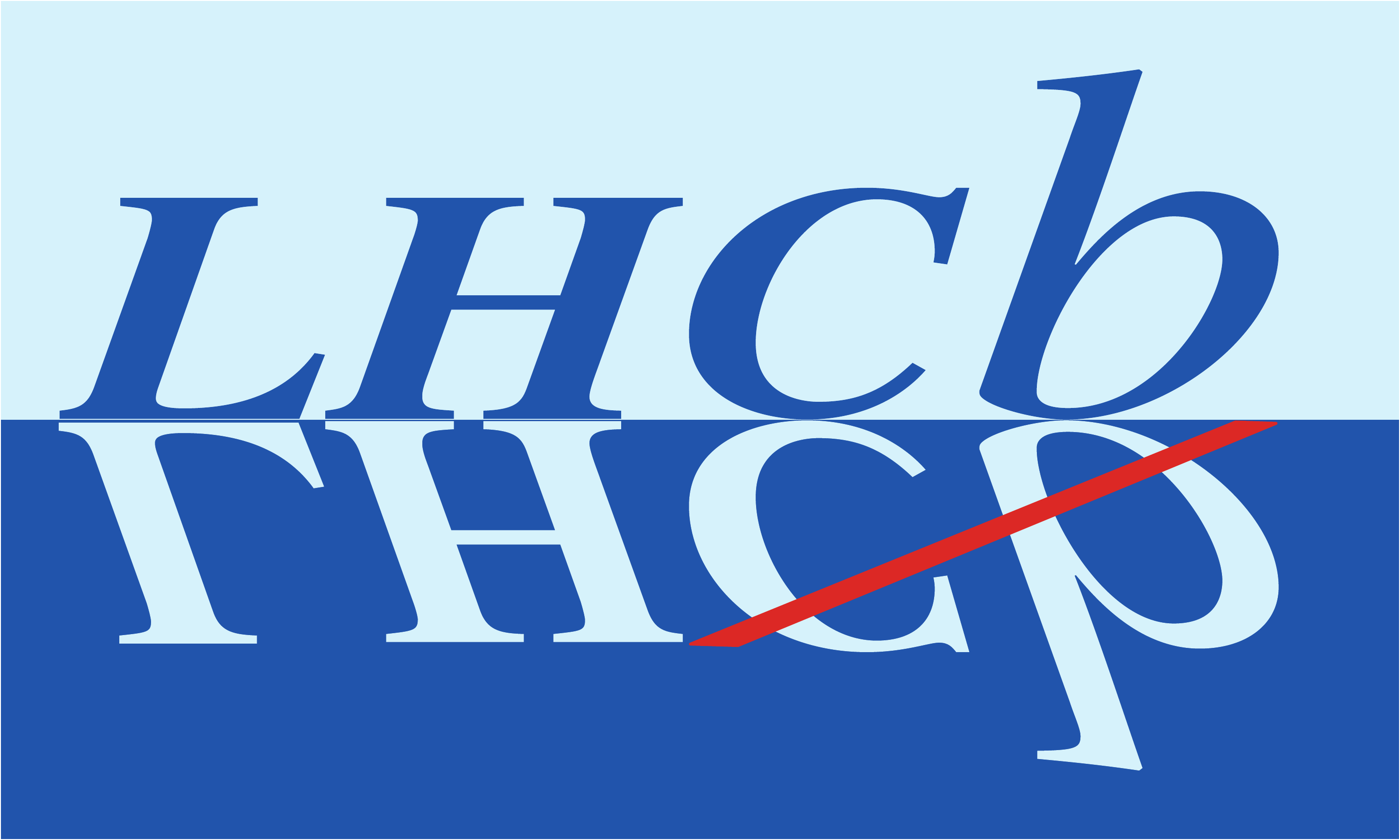}} & &}%
{\vspace*{-1.2cm}\mbox{\!\!\!\includegraphics[width=.12\textwidth]{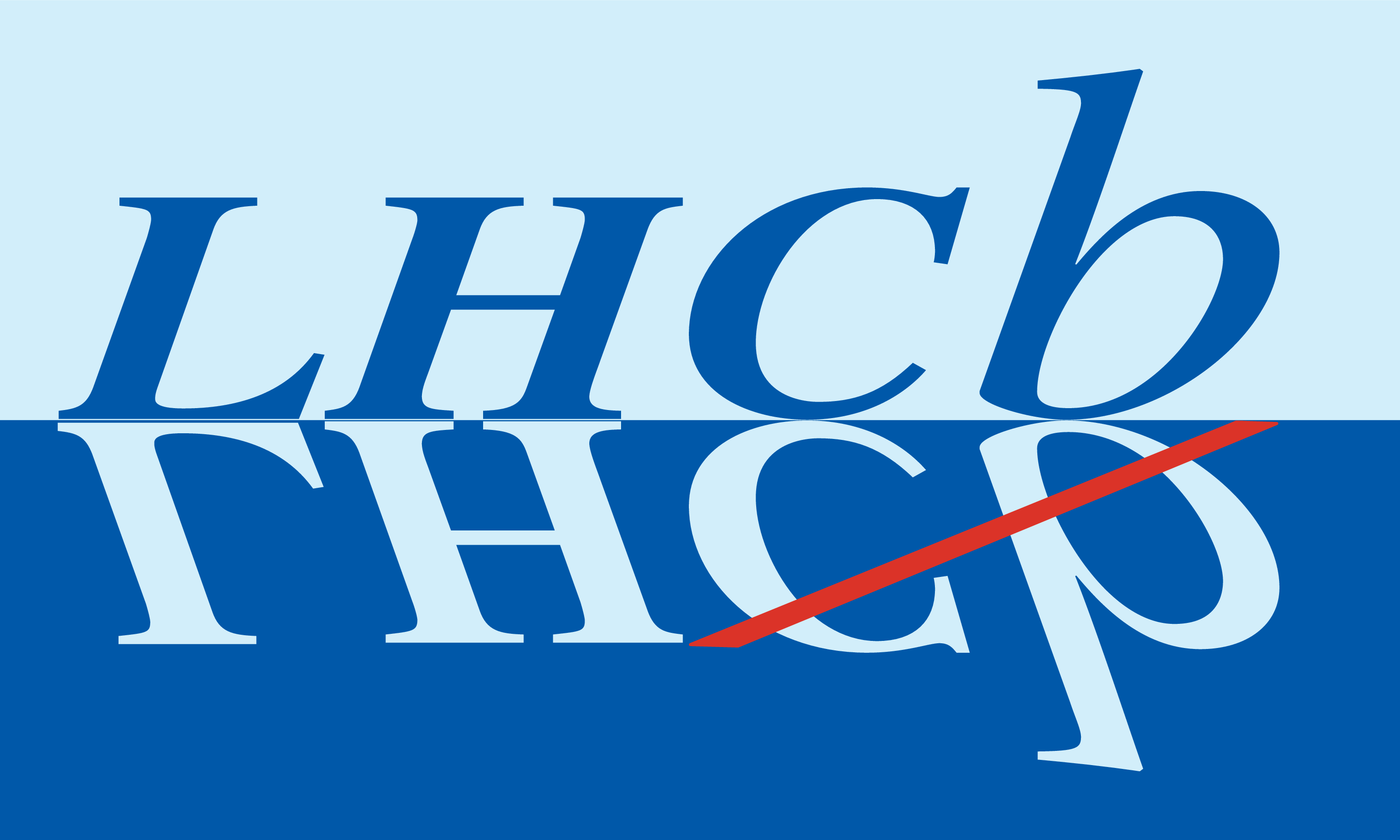}} & &}%
\\
 & & CERN-EP-2026-073 \\  % ID 
 & & LHCb-PAPER-2025-072 \\  % ID 
 & & July 13, 2026 \\ % Date - Can also hardwire e.g.: 23 March 2010
 & & \\
% not in paper \hline
\end{tabular*}

\vspace*{4.0cm}

% Title --------------------------------------------------
{\normalfont\bfseries\boldmath\huge
\begin{center}
% DO NOT EDIT HERE. Instead edit macro in main.tex to keep metadata correct
  \papertitle 
\end{center}
}

\vspace*{2.0cm}

% Authors -------------------------------------------------
\begin{center}
%In the footnote, replace 'paper' by 'Letter' in case of submission to PRL or PLB 
% Edit macro in main.tex to keep metadata correct
\paperauthors\footnote{Authors are listed at the end of this paper.}
\end{center}

\vspace{\fill}

% Abstract -----------------------------------------------
\begin{abstract}
  \noindent
  The decay \decay{\Bz}{\Lc \Lcbar \KS} is studied at LHCb for the first time using proton-proton collision data recorded by the $\lhcb$ experiment at a center-of-mass energy of $\sqrt{s}= 13\tev$, corresponding to an integrated luminosity of 5.4$\invfb$. The branching ratio relative to the decay $\Bp \to \Lc \Lcbar \Kp$ is measured to be 
  \begin{equation*}
  \frac{\BF(\Bz \to \Lc \Lcbar \KS)}{\BF(\Bp \to \Lc \Lcbar \Kp)} = 0.53 \pm 0.05 \pm 0.05,
  \end{equation*}
  where the first uncertainty is statistical and the second is systematic. Evidence is found for contributions from two resonant states, $\Xic(2923)^+$ and $\Xic(2939)^+$, in the $\Lc \KS$ system. The two states show a significance of $3.9\sigma$ relative to the nonresonant hypothesis. These two $\Xicp$ states are consistent with being the isospin partners of the states observed in $\Lc \Km$ system.
  
\end{abstract}

\vspace*{2.0cm}

\begin{center}
  Published in
  Phys.~Rev.~D 114, (2026) 012012
\end{center}

\vspace{\fill}

{\footnotesize 
% Edit macro in main.tex to keep metadata correct
\centerline{\copyright~\papercopyright. \href{\paperlicenceurl}{\paperlicence}.}}
\vspace*{2mm}

\end{titlepage}

%%%%%%%%%%%%%%%%%%%%%%%%%%%%%%%%
%%%%%  EOD OF TITLE PAGE  %%%%%%
%%%%%%%%%%%%%%%%%%%%%%%%%%%%%%%%

%  empty page follows the title page ----
\newpage
\setcounter{page}{2}
\mbox{~}
%\newpage
%
%% Author List ----------------------------
%%  You need to get a new author list!
%\input{LHCb_authorlist.tex}
%
%The author list for journal publications is provided by the Membership Committee shortly after 'approval to go to paper' has been given.
%%It will be made available on the page
%%\verb!http://www.physik.uzh.ch/~strauman/forMemCo/LHCb-PAPER-XXXX-XXX/! .
%It will be sent to you by email shortly after a paper number has beens assigned.
%The author list should be included already at first circulation, 
%to allow new members of the collaboration to verify whether they have been included correctly.
%Occasionally a misspelled name is corrected or associated institutions become full members.
%In that case, a new author list will be sent to you.
%In case line numbering doesn't work well after including the authorlist, try moving the \verb!\bigskip! after the last author to a separate line.
%
%
%The authorship for Conference Reports should be ``The LHCb
%  collaboration'', with a footnote giving the name(s) of the contact
%  author(s), but without the full list of collaboration names.

%\twocolumn
% %%%%%%%%%%%%% ---------

\renewcommand{\thefootnote}{\arabic{footnote}}
\setcounter{footnote}{0}

%%%%%%%%%%%%%%%%%%%%%%%%%%%%%%%%
%%%%%  Table of Content   %%%%%%
%%%%%%%%%%%%%%%%%%%%%%%%%%%%%%%%
%%%% Uncomment if desired
%\tableofcontents

\cleardoublepage

%%%%%%%%%%%%%%%%%%%%%%%%%
%%%%% Main text %%%%%%%%%
%%%%%%%%%%%%%%%%%%%%%%%%%

\pagestyle{plain} % restore page numbers for the main text
\setcounter{page}{1}
\pagenumbering{arabic}

%% Uncomment during review phase. 
%% Comment before a final submission.
% \linenumbers

%% This is the main body
%% It is useful to have a single file so comments are not missed in overleaf.
% \input{body}
\section{Introduction}
\label{sec:Introduction}

Multibody $B$-meson decays exhibit a rich and complex phenomenology, much of which remains unexplored. The Large Hadron Collider (LHC), with its high production rate of heavy-flavor hadrons, provides an unprecedented opportunity to deepen the understanding of $B$ meson decays and to probe both strong and weak interactions with great precision. 
Final states involving two open-charm hadrons and a strange meson, arising from the \decay{\bquark}{\cquark \cquarkbar \squark} quark-level transitions, constitute a significant fraction of the total $B$ meson decay width~\cite{PDG2024}.  These decays have proven to be particularly valuable in the discovery of both open- and hidden-charm exotic states, offering unique insights into QCD dynamics~\cite{LHCb-PAPER-2020-024,LHCb-PAPER-2020-025,LHCb-PAPER-2022-018,LHCb-PAPER-2022-019,LHCb-PAPER-2022-026,LHCb-PAPER-2022-027,LHCb-PAPER-2023-047}.

The study of the \decay{\B}{\Lc \Lcbar K} processes started with the observation of a structure at 2930\mevcc in the $\Lc K^{-}$ spectrum in \decay{\Bp}{\Lc \Lcbar \Kp} decays\footnote{Charge conjugate processes are implied throughout unless stated otherwise.} by the \babar collaboration~\cite{Aubert:2007eb}, later confirmed by the Belle collaboration~\cite{Li:2017uvv}. This structure was subsequently studied using prompt $\Lc \Km$ production in $pp$ (proton-proton) collisions~\cite{LHCb-PAPER-2020-004} and was resolved into two narrow states, namely $\Xic(2923)^0$ and $\Xic(2939)^0$. Further analysis of the \decay{\Bp}{\Lc \Lcbar \Kp} decay at LHCb confirmed the existence of these two states in $B$ decays~\cite{LHCb-PAPER-2022-028}, along with evidence for the existence of the additional state $\Xic(2880)^0$. 

Moreover, a recent LHCb study of the prompt $\Xicp\pim\pip$ production led to the discovery of the $\Xic(2923)^+$, along confirmation of $\Xic(2815)^+$, $\Xic(2970)^+$, and $\Xic(3080)^+$ states~\cite{LHCb-PAPER-2024-055}. These states, denoted as $\ensuremath{\Xires^{**}_\cquark}$, represent excited states of $\Xic$ baryon, with their quantum numbers yet to be confirmed.

\begin{figure}[bp]
    \centering 
    \includegraphics[width=0.482\textwidth]{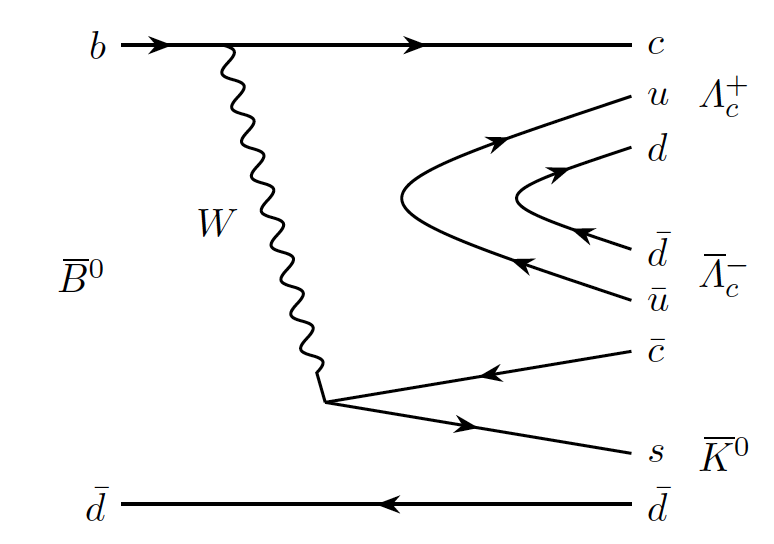}
    \includegraphics[width=0.465\textwidth]{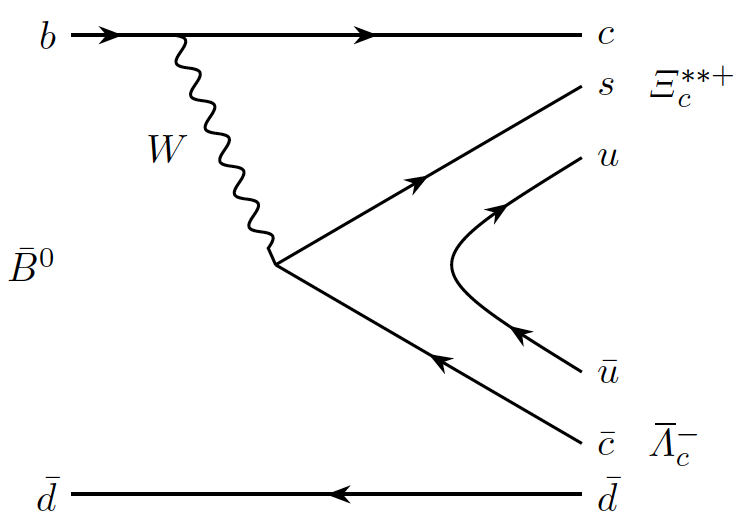}
    \caption{Feynman diagrams for the \decay{\Bz}{\Lc \Lcbar \KS} decay through (left) nonresonant and (right) \decay{\Bz}{\Lcbar \ensuremath{\Xires^{**+}_\cquark}} processes.}
    \label{fig:intro-feyn}
\end{figure}

The isospin-partner channel \decay{\Bz}{\Lc \Lcbar \KS} helps to form a more complete picture. The Belle collaboration reported a $3.9\sigma$ evidence of the charged $\Xic(2930)^+$ state in the $\Lc \KS$ spectrum~\cite{Li:2018fmq}, suggesting similar resonant structures in both \decay{\Bp}{\Lc \Lcbar \Kp}  and \decay{\Bz}{\Lc \Lcbar \KS} decays. 
The corresponding Feynman diagrams of both nonresonant and excited-state contributions are illustrated in  Fig.~\ref{fig:intro-feyn}. The unique capabilities of the LHCb detector, combined with the large $B$-meson production rates in $pp$ collisions, provide an ideal environment for detailed studies of these $\Xic$ states.

Besides open-charm systems, charmonium-like exotic states decaying to $\Lc \Lcbar$ remain intriguing. The Belle collaboration observed a near-threshold enhancement in \epem collisions with initial-state radiation~\cite{Pakhlova:2008vn}, later measured precisely by the \besiii collaboration~\cite{Ablikim:2017lct}. The recent study from \besiii does not support the resonant structure previously suggested by Belle\cite{BESIII:2023rwv}. Studies of baryon-pair systems in $B$-meson decays will provide further insight into this topic.

This paper describes a study of the \decay{\Bz}{\Lc \Lcbar \KS} decay, using $pp$ collision data at a center-of-mass energy of $\sqrt{s} = 13$ TeV collected by the LHCb detector, corresponding to an integrated luminosity of 5.4\invfb. The main goals are to measure the branching fraction relative to the \decay{\Bp}{\Lc \Lcbar \Kp} decay and to investigate potential $\Xicp$ excited states in the $\Lc \KS$ spectrum in search of the isospin partners of established $\ensuremath{\Xires^{**0}_\cquark}$ states. Moreover, the analysis also enables a search of possible charmonium-like exotic states decaying to $\Lc \Lcbar$.

%%%%%%%%%%%%%%%%%%%%%%%%%%%%%%%%%%%%%%%%%%%%%%%%%%%%%%%%%%%%%%%%%%%%%%%%%%%%%%%%%%%%%%%%%%%%%%%%%
\section{Detector and simulation}
\label{sec.Detector and simulation}

The LHCb detector~\cite{LHCb-DP-2008-001, LHCB-DP-2014-002} is a single-arm forward spectrometer covering the pseudorapidity range $2 < \eta < 5$, specifically designed for studying $b$- and $c$-hadrons. Its tracking system consists of a silicon-strip vertex detector surrounding the $pp$ interaction region, a large-area silicon-strip detector located
upstream of a dipole magnet with a bending power of about
$4 {\mathrm{\,T\,m}}$, and three stations of silicon-strip detectors and straw drift tubes placed downstream of the magnet. The tracking system provides a measurement of the momentum, \ptot, of charged particles with
a relative uncertainty that varies from 0.5\% at low momentum to 1.0\% at 200\gevc. The momentum scale is calibrated using \decay{J/\psi}{\mu^+\mu^-} and \decay{\Bp}{ J/\psi K^+} decays~\cite{LHCb-PAPER-2012-048,LHCb-PAPER-2013-011} with a relative accuracy of $3\times10^{-4}$.
The minimum distance of a track to a primary $pp$ collision vertex (PV), the impact parameter (IP), 
is measured with a resolution of $(15+29/\pt)\mum$,
where \pt is the component of the momentum transverse to the beam, in\,\gevc.
Different types of charged hadrons are distinguished using information
from two ring-imaging Cherenkov detectors. 
Photons, electrons and hadrons are identified by a calorimeter system consisting of
scintillating-pad and preshower detectors, an electromagnetic
and a hadronic calorimeter. Muons are identified by a
system composed of alternating layers of iron and multiwire
proportional chambers.
The online event selection is performed by a trigger, 
which consists of a hardware stage, based on information from the calorimeter and muon
systems, followed by a software stage, which applies a full event
reconstruction. At the hardware trigger stage, events are required to have a muon with high \pt or a
hadron, photon or electron with high transverse energy in the calorimeters. The software trigger requires a two-, three- or four-track secondary vertex with a significant displacement from any primary $pp$ interaction vertex.
Data passing the trigger requirements undergo a centralized, offline processing step,
to deliver physics-analysis-ready data across the entire \lhcb physics program~\cite{Stripping}.

Simulation is required to model the effects of the detector acceptance and the imposed selection requirements. In the simulation, $pp$ collisions are generated using \pythia~\cite{Sjostrand:2007gs,*Sjostrand:2006za} 
with a specific \lhcb configuration~\cite{LHCb-PROC-2010-056}. Decays of unstable particles are described by \evtgen~\cite{Lange:2001uf}, in which final-state radiation is generated using \photos~\cite{davidson2015photos}. The interaction of the generated particles with the detector, and its response, are implemented using the \geant toolkit~\cite{Allison:2006ve, *Agostinelli:2002hh} as described in Ref.~\cite{LHCb-PROC-2011-006}. The reconstructed signal decays in the simulation are
required to be matched with the generated signal decays. To account for known imperfections in modeling the particle identification (PID) performance, the simulation is corrected using information from control samples in data~\cite{LHCb-DP-2018-001}. The simulation samples are weighted to compensate for known differences with respect to the data in $B$ production kinematics and track multiplicity, as well as trigger efficiencies. They are used to optimize selection criteria, model signal shapes, and compute efficiencies for branching fraction measurements.

%%%%%%%%%%%%%%%%%%%%%%%%%%%%%%%%%%%%%%%%%%%%%%%%%%%%%%%%%%%%%%%%%%%%%%%%%%%%%%%%%%%%%%%%%%%%%%%%%
\section{Selection}
\label{sec.Selection}

The reconstruction of $\Bz$ candidates combines two oppositely charged $\myLc$ baryons,\footnote{The notation $\myLc$ is used to represent both \Lc and \Lcbar.} reconstructed through the \decay{\Lc}{\proton \Km \pip} decay, with one $\KS$ meson. Final-state particles from $\myLc$ decays are selected with $\pt > 100\mevc$  and required not to originate from any of the PVs. 
The $\myLc$ candidates are required to have a reconstructed mass within $\pm 110\mevcc$  of the known mass~\cite{PDG2024}. Furthermore, they are required to have $\pt > 1\gevc$, good vertex-fit quality, and the three $\myLc$ decay products tracks must have a maximum separation of 0.5 mm. At least one decay product from each $\myLc$ must have $\pt > 500 \mevc$  and $p > 5 \gevc$.

Candidate \decay{\KS}{\pip\pim} decays are reconstructed in two different categories:
the first involving \KS mesons that decay early enough within the vertex detector such that the two charged pions are reconstructed using the full tracking system, and the second containing \KS mesons whose decay vertex is such that the pions can only be reconstructed with information from the tracking detectors placed downstream of the vertex detector. The pion tracks of the two categories are referred to as \textit{long} and \textit{downstream} tracks, respectively. 
The $\KS$ candidates that use long-track pairs (LL) or downstream-track pairs (DD) are treated separately in the event reconstruction chain, with LL candidates offering superior mass and momentum resolution and DD candidates providing higher yields. The \KS candidates are selected requiring the invariant mass of the pion pair in the range $[465, 530]\mevcc$ for DD candidates and $[478, 517]\mevcc$ for LL candidates, corresponding to five times the $\KS$ mass resolution in each category. 
The $\KS$ candidates are required to have a decay vertex that is significantly displaced from the primary vertex.

For the selection of the control channel \decay{\Bp}{\Lc \Lcbar \Kp}, the $\Kp$ candidates are required to have $\pt > 500 \mevc$ and $p > 5 \gevc$. 
The \decay{\Bp}{\Lc \Lcbar \Kp} and \decay{\Bz}{\Lc \Lcbar \KS} candidates are required to have a decay vertex of good quality, be consistent with being produced at the PV and have decay time exceeding $0.2\ps$.
In addition, the final-states particles
are required to satisfy supplementary PID, and mass-windows are applied to the $B$-meson and $\myLc$ candidates. These selections suppress combinatorial background while preserving more than $95\%$ of the signal efficiency.

In order to further reduce combinatorial background, a boosted decision tree~(BDT) classifier~\cite{Breiman,AdaBoost}, implemented in the TMVA toolkit~\cite{Hocker:2007ht,*TMVA4}, is used. The classifier is trained using candidate $B$ decays from the high-mass sideband region $[5400, 5800]\mevcc$ in data as a background proxy, and simulated candidates as a signal proxy. The BDT incorporates several discriminating variables including those related to the $B$ decay topology, PID information, and kinematic properties. The BDT training is performed independently for LL or DD reconstruction categories. The optimal BDT response threshold is determined by maximizing the estimated signal significance. Separate BDTs are optimized for the signal and control channels, with nearly identical selection criteria.

%%%%%%%%%%%%%%%%%%%%%%%%%%%%%%%%%%%%%%%%%%%%%%%%%%%%%%%%%%%%%%%%%%%%%%%%%%%%%%%%%%%%%%%%%%%%%%%%%
\section{Signal yield determination}
\label{sec:Signal yield determination}

The yields of signal and background contributions are determined via three-dimensional (3D) extended unbinned maximum-likelihood fits to the invariant masses of the reconstructed $\Bz$ and $\myLc$ candidates.
For each invariant-mass distribution in $M(\Lc\Lcbar K)$, $M(\proton \Km \pip)$ and $M(\antiproton \Kp \pim)$, the signal mass shape (labeled as $S$)  is modeled by a double-sided Crystal Ball function~\cite{Skwarnicki:1986xj}, and the combinatorial background shape (labeled as $B$) is parameterised by a first-order polynomial function. A 3D fit is performed for both the signal and the control samples. The fit model is constructed as a sum of multiple terms, each of which is a direct product of one-dimensional functions representing each of the three mass distributions. 
Therefore, the function describing the signal is defined as $S_{\Bz}S_{\Lc}S_{\Lcbar}$ while the background candidates can be divided into seven different categories:
\begin{itemize}
\item Categories 1-3 (Backgrounds containing genuine $\Bz$ decays): $S_{\Bz}B_{\Lc}S_{\Lcbar}$, $S_{\Bz}S_{\Lc}B_{\Lcbar}$, $S_{\Bz}B_{\Lc}B_{\Lcbar}$;
\item Categories 4-6 (Non-$\Bz$ decays containing true $\myLc$ decays): $B_{\Bz}S_{\Lc}S_{\Lcbar}$, $B_{\Bz}B_{\Lc}S_{\Lcbar}$, $B_{\Bz}S_{\Lc}B_{\Lcbar}$;
\item Category  7 (Pure combinatorial background): $B_{\Bz}B_{\Lc}B_{\Lcbar}$.
\end{itemize}

\begin{figure}[htbp]
    \centering
        \includegraphics[width=0.32\textwidth]{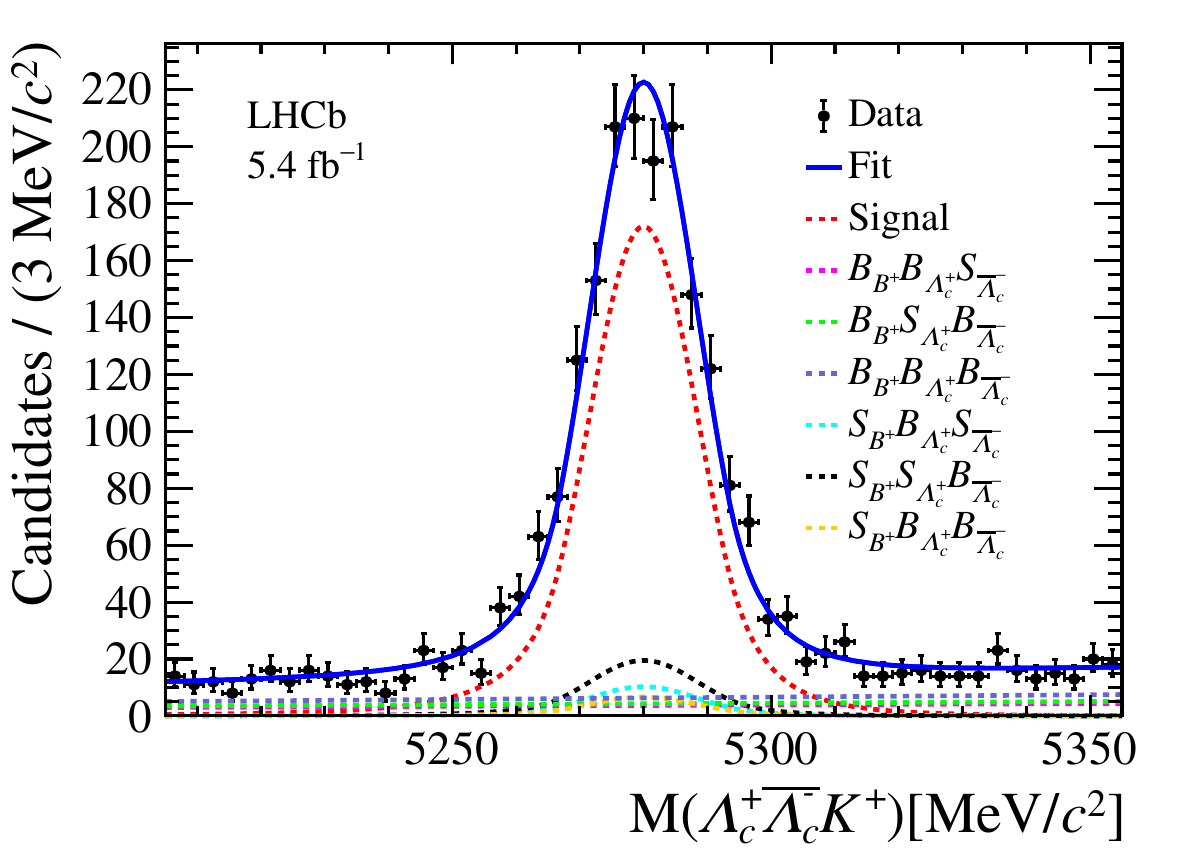}
        \includegraphics[width=0.32\textwidth]{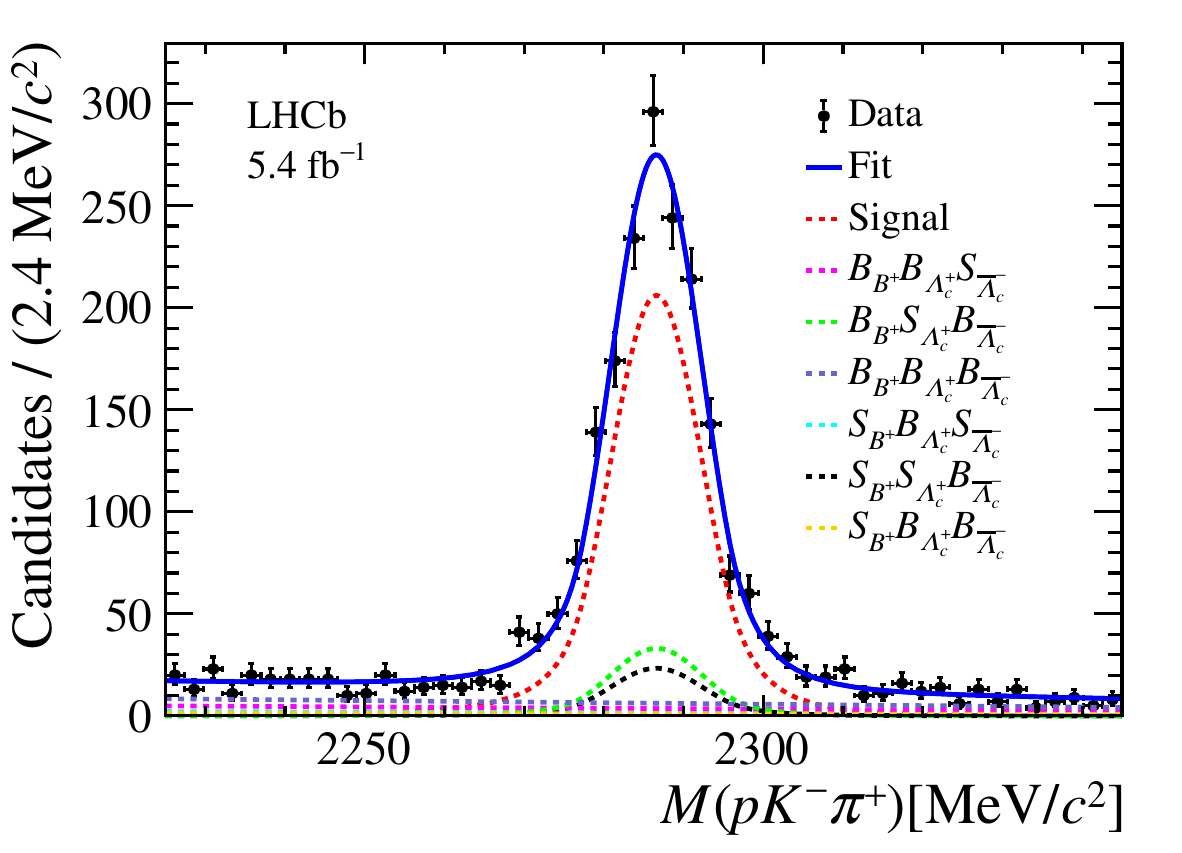}
        \includegraphics[width=0.32\textwidth]{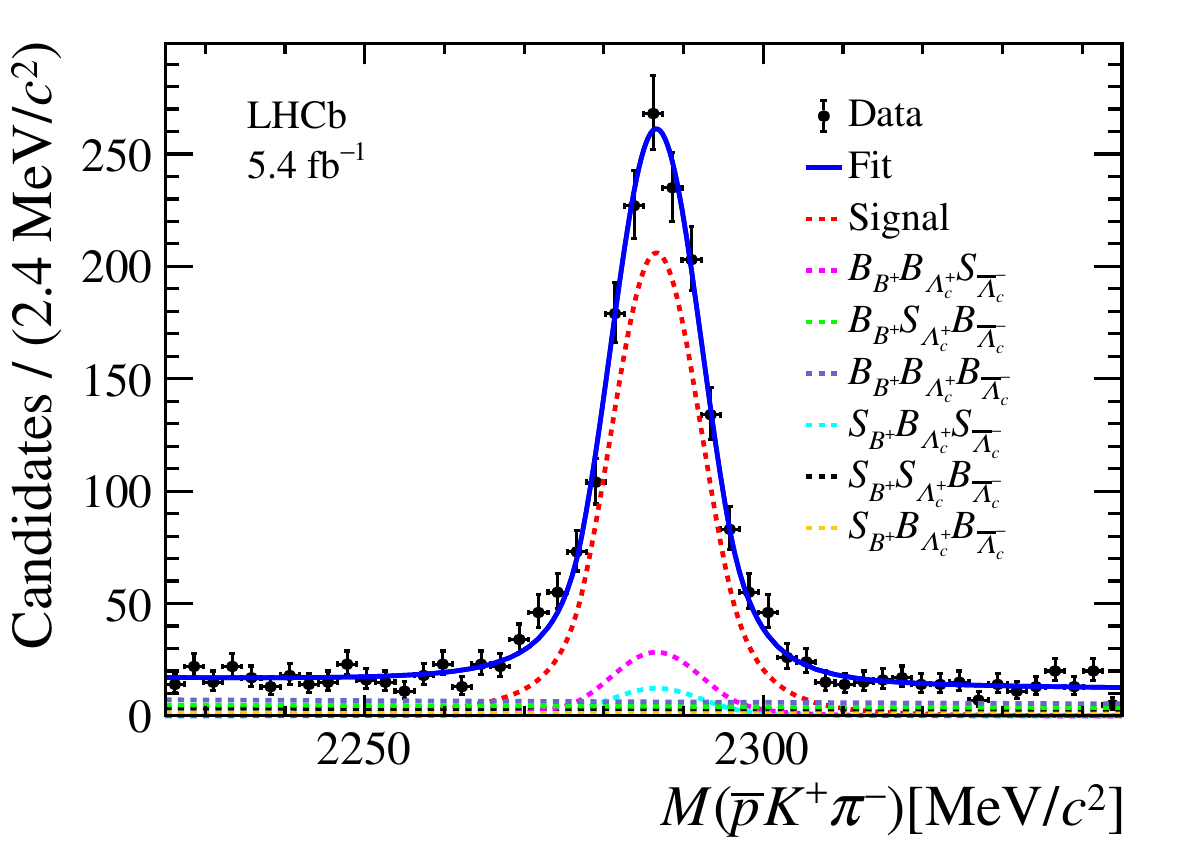}\\
        \includegraphics[width=0.32\textwidth]{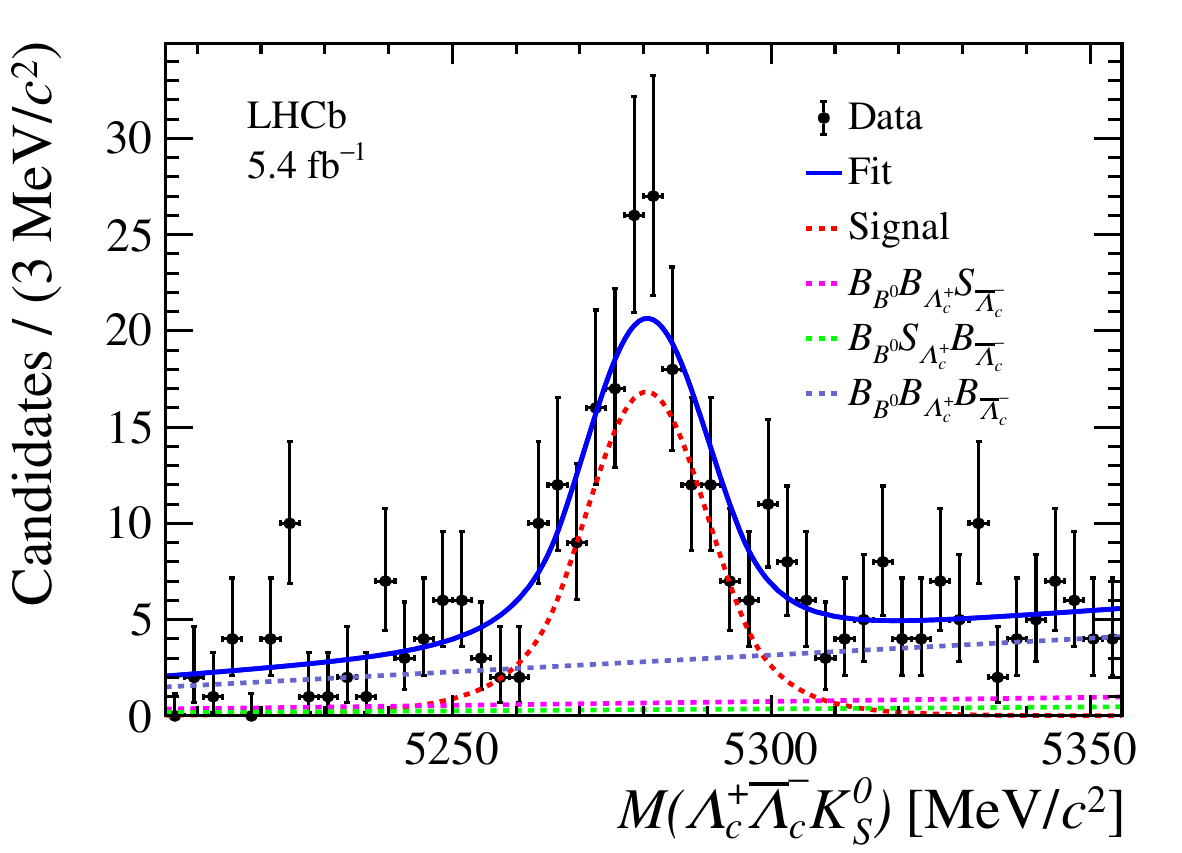}
        \includegraphics[width=0.32\textwidth]{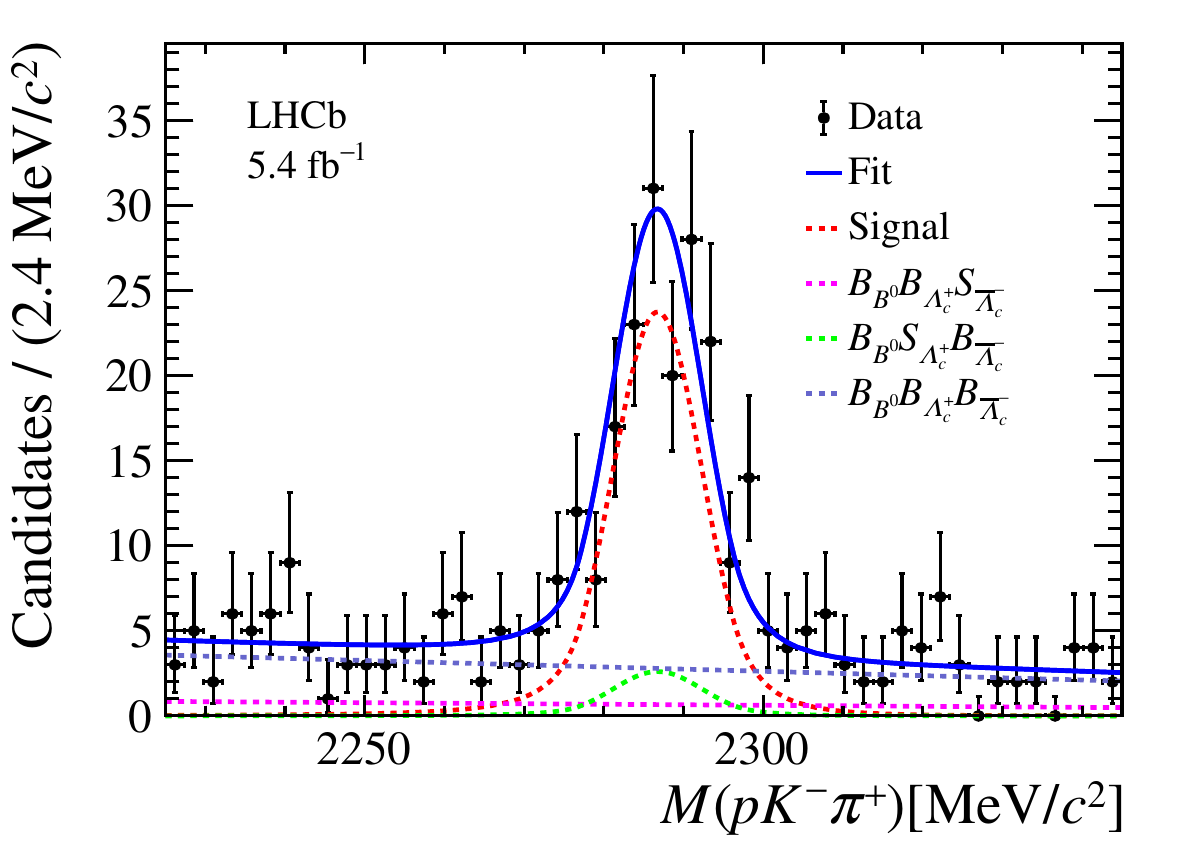}
        \includegraphics[width=0.32\textwidth]{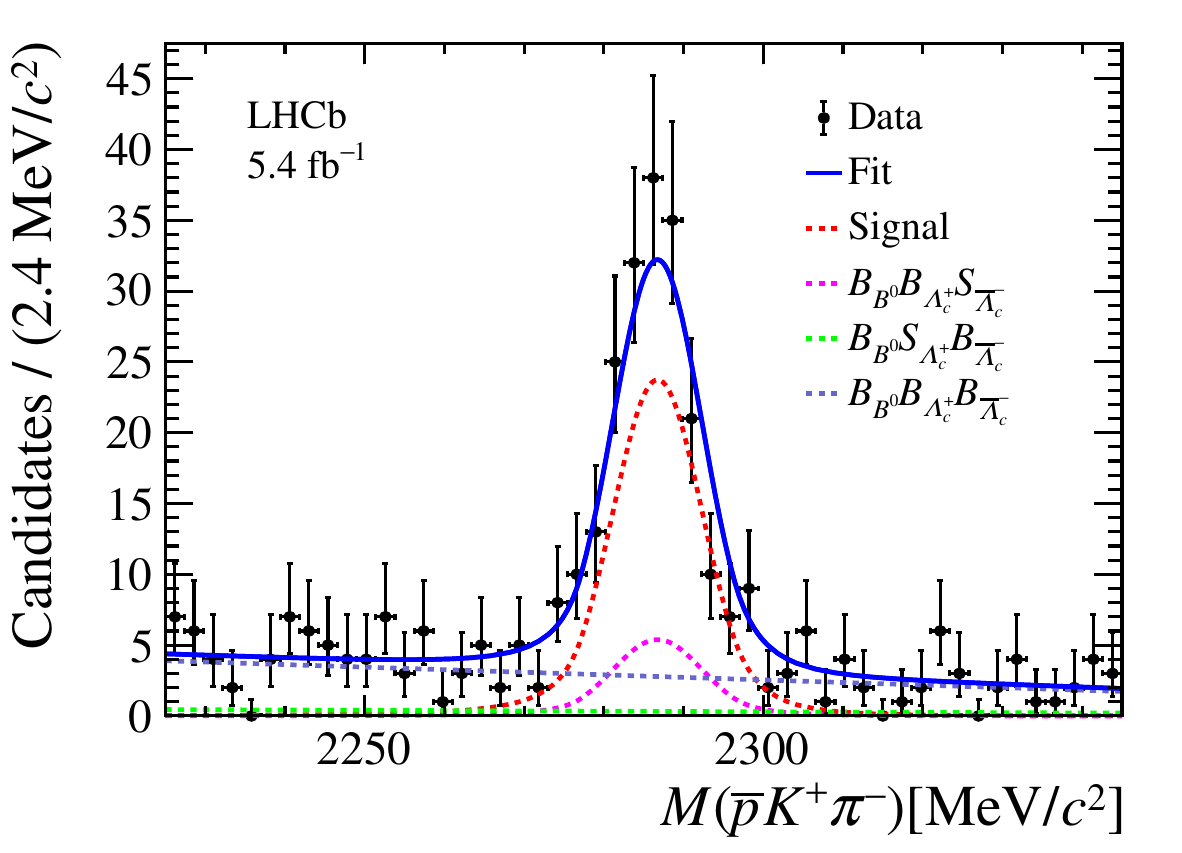}\\
        \includegraphics[width=0.32\textwidth]{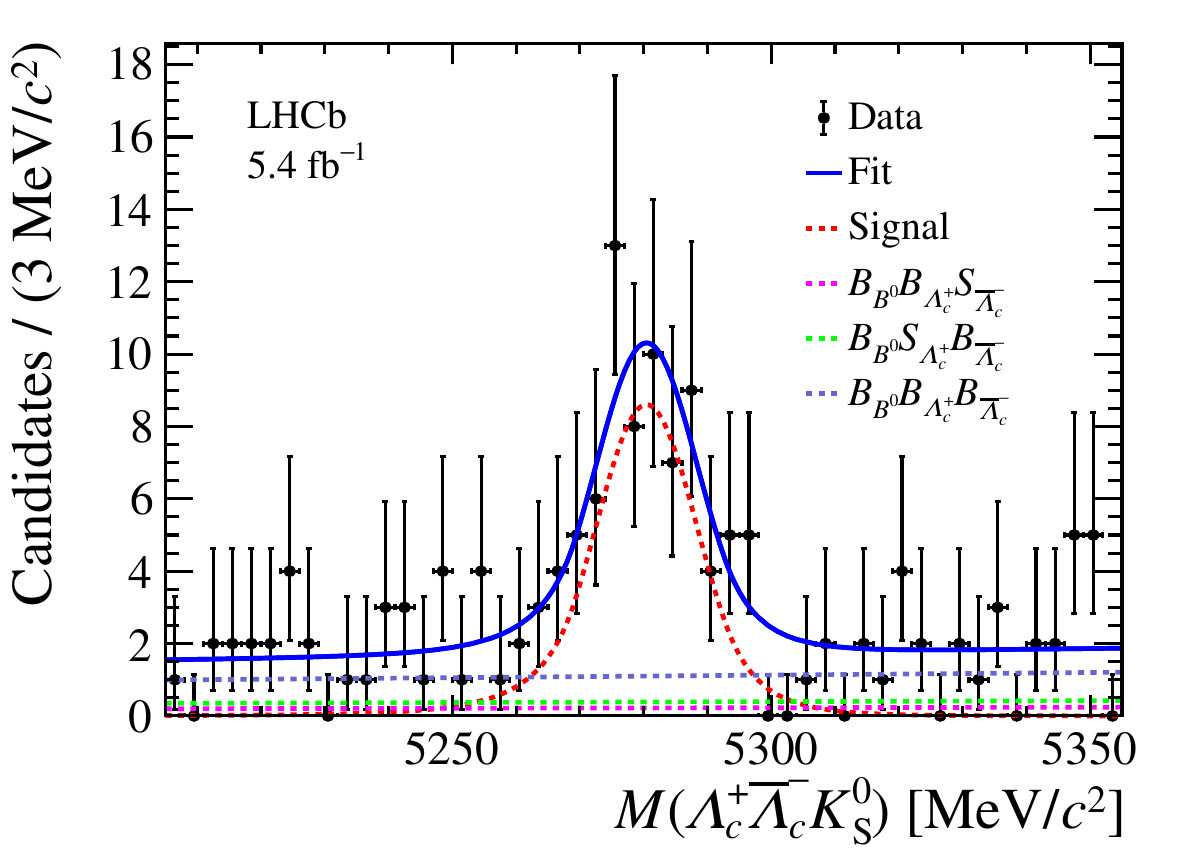}
        \includegraphics[width=0.32\textwidth]{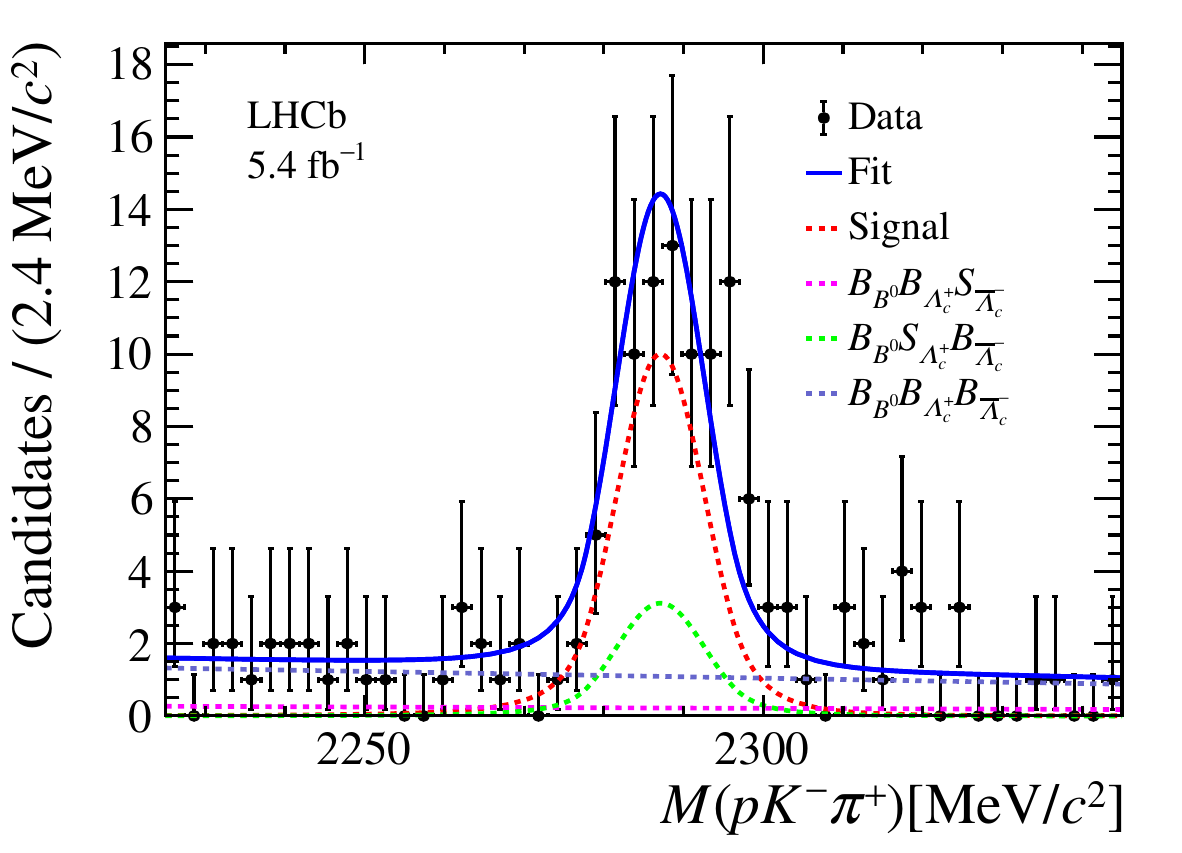}
        \includegraphics[width=0.32\textwidth]{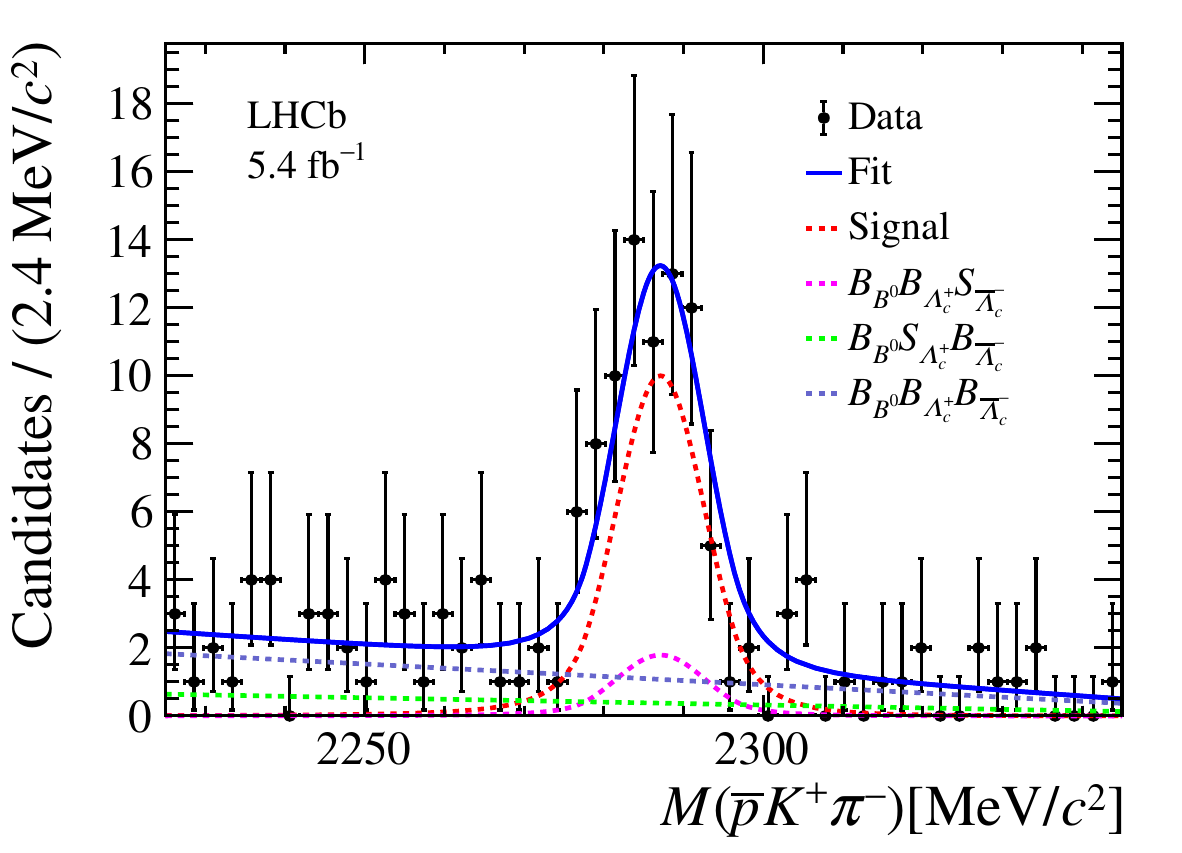}
    \caption{\enspace Fits to the (top row) \decay{\Bm}{\Lc \Lcbar \Km}, (middle row) \decay{\Bz}{\Lc \Lcbar \KS} DD and (bottom row) LL categories. The projections along $M(\Lc\Lcbar K)$, $M(\proton \Km \pip)$ and $M(\antiproton \Kp \pim)$ are shown in left, middle and right columns, respectively.}
    \label{fig:LcLcKs_combined_DDLL_LcLcK}
\end{figure}

It is found that the backgrounds in categories 1-4 are negligible for the signal mode, while this is not the case for categories 5-7. All the tail parameters of the Crystal Ball functions are fixed to the values obtained from simulation and shared between the \Lc and \Lcbar baryons. A similar fit strategy is used for the control mode. Separate fits are performed for LL and DD categories. The invariant-mass distributions of the selected \Bz and $\myLc$ candidates after applying all the aforementioned selection requirements are shown in Fig.~\ref{fig:LcLcKs_combined_DDLL_LcLcK}. The signal yields are determined to be $148\pm 14$ and $62\pm 9$ for DD and LL modes, respectively. The yields in the DD and LL modes are consistent with each other accounting for differences in efficiencies. The yield of the \decay{\Bp}{\Lc \Lcbar \Kp} decay is~$1335\pm 43$.

%%%%%%%%%%%%%%%%%%%%%%%%%%%%%%%%%%%%%%%%%%%%%%%%%%%%%%%%%%%%%%%%%%%%%%%%%%%%%%%%%%%%%%%%%%%%%%%%%
\section{Branching fraction}
\label{sec:Branching fraction}

The ratio of branching fractions is defined as:
\begin{equation}
\begin{split}
%\mathcal{R} \equiv 
&\frac{\BF(\Bz \to \Lc \Lcbar \KS)}{\BF(\Bp \to \Lc \Lcbar \Kp)} = \\
&  \frac{N(\Bz \to \Lc \Lcbar \KS)}{N(\Bp \to \Lc \Lcbar \Kp)}
\times \frac{\eps(\Bp \to \Lc \Lcbar \Kp)}{\eps(\Bz \to \Lc \Lcbar \KS)} 
\times \frac{1}{\BF(\KS \to \pip \pim)} \times \frac{f_u}{f_d},
\end{split}
\end{equation}
where $N$, \eps, \BF and ${f_u}/{f_d}$ denote the observed signal yields, the total experimental efficiencies, the branching fraction, and the ratio of hadronization fractions of \bquark quarks to \Bp and \Bz mesons, respectively. The branching fraction $\BF(\KS \to \pip \pim) = (69.20 \pm 0.05)\%$ is taken from Ref.~\cite{PDG2024}. The ratio ${f_u}/{f_d}$ is assumed to be equal to unity~\cite{Averages_properties}.
The yields of the signal and control modes are obtained from the 3D fit described in Sec.~\ref{sec:Signal yield determination}, applied to candidates satisfying additional trigger requirements to reduce trigger-related uncertainties. The corresponding yields of signal and control modes are $128 \pm 12$ and $971 \pm 36$, respectively.
The total efficiencies are determined from simulation and defined as the product of efficiencies corresponding to detector acceptance, reconstruction,  offline preselection, trigger and multivariate selection. Simulated samples are weighted to ensure distributions in the key variables match those in data. A data-driven correction for tracking efficiency is applied. The tracking efficiency for charged tracks is estimated using data, and the difference between simulation and data is obtained from a correction table as a function of the momentum of the track, its pseudorapidity, and the track multiplicity of the event. The total efficiencies of the signal and control decays are $(1.21 \pm 0.01)\times 10^{-4}$ and $(3.36 \pm 0.02)\times 10^{-4}$, respectively, where the uncertainties are statistical only.

The systematic uncertainties on the relative branching fraction measurement, apart from external inputs, are related to the signal yields and efficiencies. Due to the similar topology of the signal and control decays, many sources of systematic uncertainties  either cancel or are largely suppressed in the ratio of branching fractions. The remaining systematic uncertainties are outlined below and summarized in Table~\ref{table:summary_sys_BR}.

\begin{table}[tbp]
\centering
\caption{Sources of systematic uncertainties on the branching ratio measurement.}
\begin{tabular}{l c|c}
\hline
\multicolumn{2}{c|}{Source}    & Relative uncertainty [\%] \\
\hline
\multicolumn{2}{c|}{Signal model}   & 3.3\\
\multicolumn{2}{c|}{Mass correlation} 
&2.9\\
\multicolumn{2}{c|}{Fit assumption}  &1.9\\
&Fit range   &1.2\\
\multicolumn{2}{c|}{Simulation sample size}  &3.2\\
\multicolumn{2}{c|}{Weighting}   &6.2\\
\multicolumn{2}{c|}{Tracking efficiency}  &0.8\\
\multicolumn{2}{c|}{Hadronic interaction}  & 1.7 \\
\hline
\multicolumn{2}{c|}{Total} & 8.7\\
\hline
\end{tabular}
\label{table:summary_sys_BR}
\end{table}

The determination of the signal yields in the $\Bz$ and $\Bp$ decay channels is affected by the choice of the fit models for the various components. 
For the \B meson mass distribution, the Hypatia function~\cite{Santos:2013gra} and the sum of two Gaussian functions are considered as alternative models for the signal component. For the $\myLc$ baryon mass peaks, a sum of three Gaussian functions is considered as the alternative model.
The correlation between $B$ and $\myLc$ masses is ignored in the baseline fit, and the correlation effect is studied by refitting the $B$ mass while constraining the $\myLc$ mass to its known value~\cite{PDG2024}. 
Regarding the fit assumption, the reconstructed $B$-mass distribution $(S_B)$ for backgrounds is allowed to differ from that of the signal component. Therefore,  separate signal $B$ shapes for different components are studied in alternative fits. In addition to this, the fitting ranges for both $B$ and $\myLc$ masses are varied.

The systematic uncertainties in the efficiency estimation arise from the limited size of the simulation samples and imperfect simulation of the track reconstruction and the detector response. 
The uncertainty due to weighting of the simulated samples for the track multiplicity is estimated using pseudoexperiments.
In each pseudoexperiment, the applied weight in each track multiplicity bin is varied according to a Gaussian distribution, whose mean and width are taken from the baseline correction factor and its associated uncertainty, respectively. A fit with a Gaussian function is performed to the distribution of deviations of the recalculated efficiency. The resulting width is assigned as the corresponding systematic uncertainty. 
The tracking efficiency correction is validated using the same method and the associated uncertainty is found to be negligible. In addition, the final states of the signal mode and the control mode differ by one track, hence an additional relative uncertainty for the tracking efficiency is considered. 
Possible effects from intermediate resonances are taken into account by weighting the phase space variables to match the distribution in data, and the resulting variation is assigned as the systematic uncertainty are included in the weighting systematic uncertainty.

Uncertainties of 1.1\% and 1.4\% due to hadronic interactions are assigned for kaons and pions to account for the uncertainties in material budget and interaction cross-section in simulations~\cite{DeCian:1402577} for different particle types, in both signal and control decays.

Additional sources of systematic uncertainties are considered but found negligible. The PID efficiency is obtained from calibration data in separate regions of momentum and pseudorapidity. Systematic uncertainties are assessed through pseudoexperiments and by varying the limits of the these regions. Potential background contamination in the simulation is evaluated by fitting simulated samples without matching requirements to the generated signals. All related systematic effects are found to be negligible for both signal and control channels.

%%%%%%%%%%%%%%%%%%%%%%%%%%%%%%%%%%%%%%%%%%%%%%%%%%%%%%%%%%%%%%%%%%%%%%%%%%%%%%%%%%%%%%%%%%%%%%%%%
\section{Intermediate resonances}
\label{sec:Resonant states}

\begin{figure}[bp]
    \centering
    \includegraphics[width=0.5\textwidth]{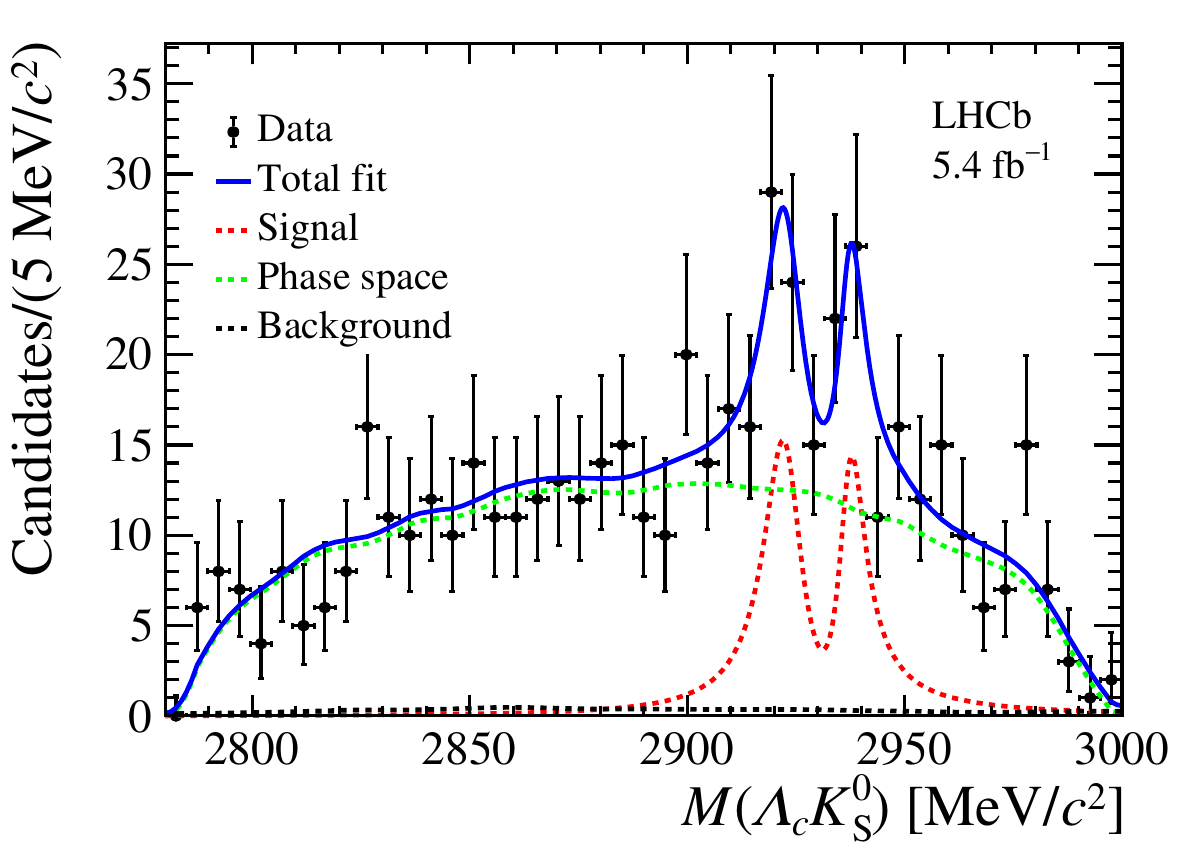}
    \caption{Fit to the $\myLc\KS$ mass spectrum considering $\Xic(2923)^+$ and $\Xic(2939)^+$ states.}
    \label{fig:LcKs_spectrum}
\end{figure}

To study the intermediate states with improved mass resolution while enhancing signal purity, stringent mass requirements are imposed on the \Bz candidates in range of $[5230, 5330] \mevcc$ and the $\myLc$ candidates in range of $[2255, 2315]\mevcc$. Subsequently kinematic quantities are recalculated with the \Lc and \KS candidates masses constrained to their established values. The signal channel \decay{\Bz}{\Lc \Lcbar \KS} is a self-conjugate decay process, where excited charged $\ensuremath{\Xires^{**+}_\cquark}$ states are expected to contribute to both the \Lc\KS and \Lcbar\KS invariant-mass spectra. The projection of the \Lc\KS and \Lcbar\KS invariant-mass spectra, denoted $\myLc\KS$, is shown in Fig.~\ref{fig:LcKs_spectrum} for the combined LL and DD samples, together with the result of an unbinned maximum-likelihood fit that includes two $\ensuremath{\Xires^{**+}_\cquark}$ states, as detailed below.

\begin{figure}[tbp]
    \centering
    \includegraphics[width=0.5\textwidth]{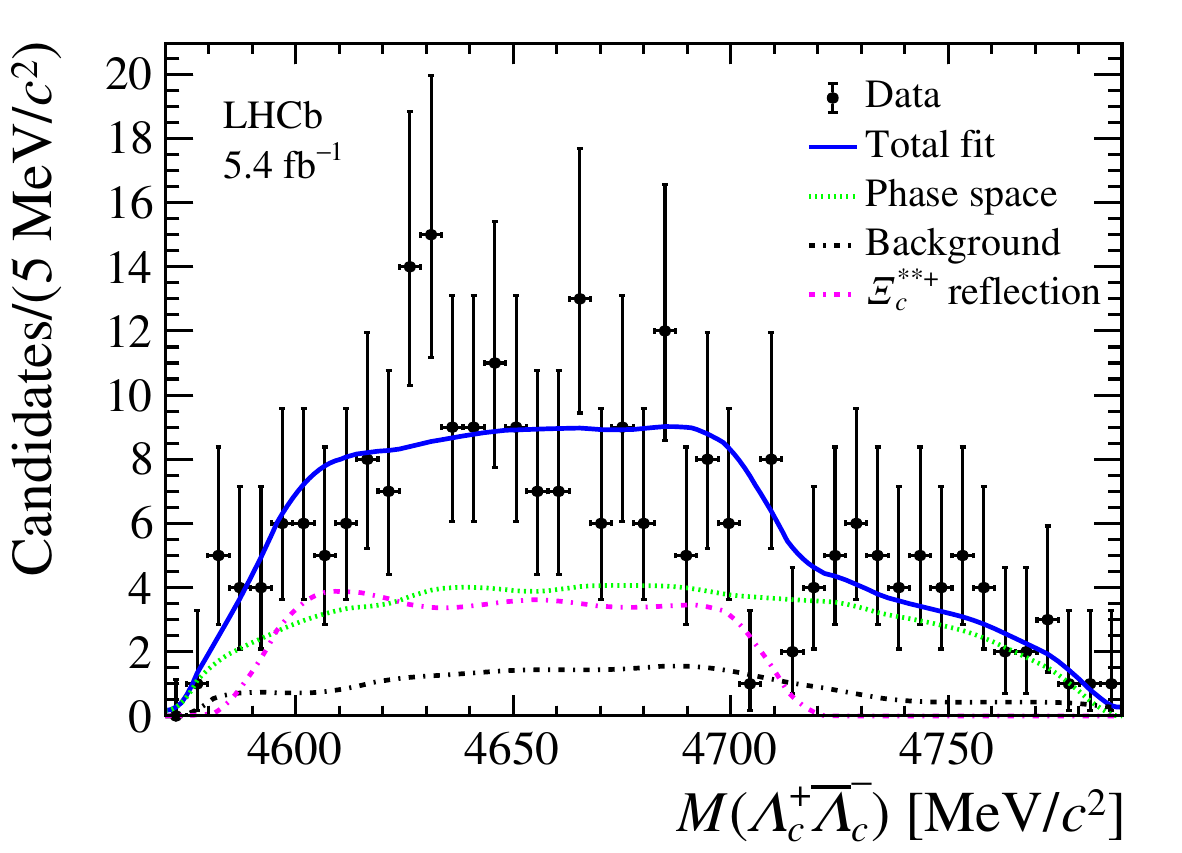}
    \caption{Fit to the $\Lc \Lcbar$ mass spectrum.}
    \label{fig:LcLc_spectrum_nopeak}
\end{figure}

The \Lc\Lcbar invariant-mass spectra are shown in Fig.~\ref{fig:LcLc_spectrum_nopeak}. The fit components are similar to those used in the fit of $\myLc\KS$ combination. In this case, a contribution due to the reflection of the $\ensuremath{\Xires^{**+}_\cquark}$ states is included, with the shape determined from a phase-space generator-level simulation of \decay{\Bz}{\Lcbar\ensuremath{\Xires^{**+}_\cquark}} decays. No significant structure emerges within the current statistical precision.

The fit model for the $\myLc\KS$ mass spectrum is constructed as
\begin{equation}
f(M) = g_{\ensuremath{\Xires^{**}_\cquark}} f_{\ensuremath{\Xires^{**}_\cquark}} \otimes G(0,\sigma_{\rm res}) + g_{\rm bkg} f_{\rm bkg} +(1 - g_{\ensuremath{\Xires^{**}_\cquark}} - g_{\rm bkg}) f_{\rm phsp},
\label{eq:resonance model}
\end{equation}
where $ g_{\ensuremath{\Xires^{**}_\cquark}}$ and $ g_{\rm bkg}$ are the fractional yields of $\ensuremath{\Xires^{**}_\cquark}$ and background, respectively,  leaving $(1 - g_{\ensuremath{\Xires^{**}_\cquark}} - g_{\rm bkg})$ as the fraction of the nonresonant component. The function $f_{\ensuremath{\Xires^{**}_\cquark}}$ represents the probability density function of excited $\ensuremath{\Xires^{**}_\cquark}$ states and $G(0,\sigma_{\rm res})$ is a Gaussian function parameterizing the detector resolution. Separate detector resolutions are used for the DD and LL subsamples. To combine these two subsamples, the weight factors for DD and LL are based on the statistical precision in the considered intervals. The resolution shows dependence on the variable $M(\myLc \KS)$ which is modeled by a second-order polynomial function whose parameters are determined from simulation. The background shape, $f_{\rm bkg}$, is determined from the data excluding the $\Bz$ and $\myLc$ mass regions.   The shape of the nonresonant component, $f_{\rm phsp}$, is obtained from simulation of pure phase space \decay{\Bz}{\Lc \Lcbar \KS} decays. 

In the baseline fit, the sum of $\Xic(2923)^+$ and $\Xic(2939)^+$ contributions is modeled as

\begin{equation}\label{eq:fxicstst_baseline}
f_{\ensuremath{\Xires^{**}_\cquark}} = p q \left(\left| \mathcal{M}_{\Xic(2923)^+} \right|^2+ \alpha^2 \left| \mathcal{M}_{\Xic(2939)^+}\right|^2\right) ,
\end{equation}
where $p$ is the $\ensuremath{\Xires^{**}_\cquark}$ momentum in the $\Bz$ rest frame, $q$ is the $\myLc$ (or $\KS$) momentum in the $\ensuremath{\Xires^{**}_\cquark}$ rest frame 
%, $q_r$ denotes its value at the resonance peak, 
and $\alpha$ quantifies the relative contribution of the two peaks. The amplitude $\mathcal{M}$ is parametrised using Blatt--Weisskopf barrier factors~\cite{Blatt:1952ije}, $B_L$, and a relativistic Breit--Wigner (RBW) function~\cite{LHCb-PAPER-2016-019}, defined as 
\begin{align}
\mathrm{RBW}(M) = \frac{1}{M_0^2 - M^2 - i M_0 \Gamma(M)}, \\
\Gamma(M) = \Gamma_0 \left(\frac{q}{q_0}\right)^{2L+1} \left(\frac{M_0}{M}\right) B_L^2,
\end{align}
where $q_0$ is the value of the momentum $q$ at the resonance peak, and $L$ is the orbital angular momentum between the $\myLc$ and \KS hadrons. The parameters $M_0$ and $\Gamma_0$ denote the pole mass and width, which are determined from the fit. The $B_L$ factors are evaluated assuming an effective radius of the resonant state of $3\gev^{-1}$~\cite{LHCb-PAPER-2020-004}, corresponding to approximately $0.6\fm$~\cite{LHCb-PAPER-2014-061}.

Due to the limited signal yields, the quantum numbers of the $\ensuremath{\Xires^{**}_\cquark}$ states cannot be determined. The spin-parity ($J^P$) assignments in this paper follow the same configuration as in the previous study of \decay{\Bp}{\Lc \Lcbar \Kp} decays~\cite{LHCb-PAPER-2022-028}, with the orbital angular momentum between $\myLc$ and $\KS$ set to two and the $J^P$ of both $\Xicp$ states assumed to be $\frac{3}{2}^-$~\cite{Wang_2020}. Given the $\KS$ and $\myLc$ particles have $J^P = 0^-$ and $\frac{1}{2}^+$, respectively, the $P$-parity of $\Xicp$ is consequently determined to be negative. 
This provides a baseline model, allowing for subsequent comparison with the models that incorporate interference.
Alternative hypotheses of S-wave and P-wave configurations are considered as part of the systematic uncertainty evaluation. 

The $M(\myLc \KS)$ mass distribution is shown in Fig.~\ref{fig:LcKs_spectrum}, together with the result of the baseline fit, which includes contributions from both the $\Xic(2923)^+$ and $\Xic(2939)^+$ states. Their combined yield is $86 \pm 22$, while their mass and width are reported in Table~\ref{table:fit_parameters_LcKs}. 
The significance of the two-resonance structure with respect to the background-only hypothesis is evaluated using Wilks' theorem~\cite{Cowan_2011} and found to be~$3.9\sigma$. When the null hypothesis is set to a single peak, allowing for changes in both the peak mass and its width, the significance of the two-resonance structure is reduced to $1.7\sigma$.

Systematic uncertainties on the mass and width measurements from various sources are studied. Multiple alternative assumptions on the fit parameters are tested to determine the possible uncertainties related to the choice of the signal model. The effective radius of $\Xicp$ resonances is changed to $2.0 \gev^{-1}$ and $4.0 \gev^{-1}$. Alternative spin-parity assignments of the two $\Xicp$ states, $J^P = \frac{1}{2}^+$, $\frac{1}{2}^-$, $\frac{3}{2}^+$ or $\frac{3}{2}^-$, are considered. Different resonances with the same spin-parity that are close in mass can interfere. To account for this, Eq.~\ref{eq:fxicstst_baseline} is replaced with the coherent sum 
\begin{equation}
p q\left| \mathcal{M}_{\Xic(2923)^+}+ \alpha e^{i\theta} \mathcal{M}_{\Xic(2939)^+} \right|^2,
\end{equation} 
where $\alpha$ is a free parameter in the fit. The relative phase $\theta$ is scanned in the range of $[-\pi,\pi]$ to determine a related systematic uncertainty. 
The background yields are constrained to the results of the baseline fit using Gaussian priors. The maximum variation of the fit results for each alternative model assumption described above is assigned as the corresponding uncertainty, and added in quadrature to the total systematic uncertainty.

To study possible biases on the measured mass and width of the $\Xic(2923)^+$ and $\Xic(2939)^+$ states, 1000 pseudoexperiments are performed in which all fit parameters, except the masses and widths, are fixed. Each pseudosample is generated according to the baseline fit result where the total number of signal and background candidate yields are fluctuated according to Poisson distributions.
The deviations between the mean values obtained from the pseudoexperiment results and the generated values are evaluated. The deviations are smaller than the statistical uncertainties and are considered in the systematic uncertainty determination.

The systematic uncertainty associated with the modeling of the mass-resolution parameter is evaluated using an alternative second-order polynomial, obtained by fitting the mass-resolution dependence with a different binning.
In addition, several other sources of systematic uncertainty only affect the mass measurement. These include the momentum uncertainty, evaluated by shifting the momentum scale used in the reconstruction of charged tracks by a factor of $3 \times 10^{-4}$~\cite{LHCb-PAPER-2020-004, LHCb-PAPER-2013-011} in simulated decays, and the imperfect modeling of the energy loss in the detector material~\cite{LHCb-PAPER-2010-001}. The deviation from the baseline fit is assigned as the systematic uncertainty.

\begin{table}[tbp]
\centering
\caption{Measured masses and  widths of the  $\Xic(2923)^+$ and $\Xic(2939)^+$ states.}
\begin{tabular}{c|c|c}
\hline
State  &  Mass [\mevcc\!]  & Width [\mevcc\!]\\
\hline
$\Xic(2923)^+$ & $2923.1 \pm 1.8 \pm 0.8$ & $11.6 \pm 5.2 \pm 2.6$\\
$\Xic(2939)^+$ & $2937.5 \pm 1.6 \pm 0.9$ & \phantom{1}$7.1 \pm 4.1 \pm 2.4$\\
\hline
\end{tabular}
\label{table:fit_parameters_LcKs}
\end{table}

\begin{table}[tbp]
\centering
\caption{Summary of systematic uncertainties on the masses and widths of the $\Xic(2923)^+$ and $\Xic(2939)^+$ states.} 
\begin{tabular}{c|cc|cc}
\hline
 & \multicolumn{2}{c|}{$\Xic(2923)^+$} & \multicolumn{2}{c}{$\Xic(2939)^+$} \\
Source & $M$ [\mevcc\!]  &  $\Gamma$ [\mevcc\!] & $M$ [\mevcc\!] & $\Gamma$ [\mevcc\!]\\
\hline
Model assumption & \phantom{$-$}0.75 & \phantom{$-$}1.20 & \phantom{$-$}0.82 &1.20 \\
Fit procedure & \phantom{$-$}0.38  & \phantom{$-$}2.32  & \phantom{$-$}0.37  &2.08  \\
Mass resolution & \phantom{$-$}0.00  & \phantom{$-$}0.02 & \phantom{$-$}0.08 &0.03 \\
Momentum scaling &\phantom{$-$}0.04  &\phantom{$-$}--  & \phantom{$-$}0.05  &--  \\
Energy loss & \phantom{$-$}0.04  &\phantom{$-$}--  & \phantom{$-$}0.04  &--  \\
\hline
Total & \phantom{$-$}0.84  & \phantom{$-$}2.61  & \phantom{$-$}0.91  &2.40  \\
\hline
\end{tabular}
\label{table:summary_sys_Xic}
\end{table}

A summary of the systematic uncertainties is listed in Table~\ref{table:summary_sys_Xic}. The uncertainties
from different sources are added in quadrature to give the total systematic uncertainty.

%%%%%%%%%%%%%%%%%%%%%%%%%%%%%%%%%%%%%%%%%%%%%%%%%%%%%%%%%%%%%%%%%%%%%%%%%%%%%%%%%%%%%%%%%%%%%%%%%
\section{Results and summary}
\label{sec:Results and summary}

The \decay{\Bz}{\Lc \Lcbar \KS} decay is studied at LHCb using \proton\proton collision data collected at \mbox{$\sqrt{s} = 13$ TeV}, corresponding to an integrated luminosity of 5.4\invfb. The branching ratio of the \decay{\Bz}{\Lc \Lcbar \KS} decay is measured relative to \decay{\Bp}{\Lc \Lcbar \Kp} as

\begin{equation*}
\begin{aligned}
\frac{\BF(\Bz \to \Lc \Lcbar \KS)}{\BF(\Bp \to \Lc \Lcbar \Kp)} &= 0.53 \pm 0.05 \pm 0.05,  
\end{aligned}
\end{equation*}
where the uncertainties are statistical and systematic, respectively. 
Using the \decay{\Bp}{\Lc \Lcbar \Kp} branching fraction from Ref.~\cite{PDG2024}, the branching fraction of the \decay{\Bz}{\Lc \Lcbar \KS} decay is determined to be
\begin{equation*}
\begin{aligned}
\BF (\Bz \to \Lc \Lcbar \KS) &= (2.60 \pm 0.26  \pm 0.23 \pm 0.37 ) \times 10^{-4},
\end{aligned}
\end{equation*}
where the third uncertainty is due to the external branching fraction of the \decay{\Bp}{\Lc \Lcbar \Kp} decay~\cite{PDG2024}. This result is the most precise measurement to date.

The fit to the $\myLc \KS$ mass spectrum shows evidence for the $\Xic(2923)^+$ and $\Xic(2939)^+$ resonances with a significance of 3.9$\sigma$. Both the masses and widths are consistent with measurements from the isospin-partner decay \decay{\Bp}{\Lc \Lcbar \Kp}~\cite{LHCb-PAPER-2022-028} and from prompt measurements in \decay{\Xic}{\Lc \Km} decays~\cite{LHCb-PAPER-2020-004}. No significant structure is seen in the $\Lc \Lcbar$ spectrum.

% Comment this in for paper drafts; do not include this in analysis note, conference and figure reports
% \input{LHCb/acknowledgements}
\section*{Acknowledgements}
%
% These Acknowledgements valid from 21/02/26
%
\noindent We express our gratitude to our colleagues in the CERN
accelerator departments for the excellent performance of the LHC. We
thank the technical and administrative staff at the LHCb
institutes.
We acknowledge support from CERN and from the national agencies:
ARC (Australia);
CAPES, CNPq, FAPERJ and FINEP (Brazil); 
MOST and NSFC (China); 
CNRS/IN2P3 and CEA (France);  % added CEA 26/02/2026
BMFTR, DFG and MPG (Germany);
INFN (Italy); 
NWO (Netherlands); 
MNiSW and NCN (Poland); 
MCID/IFA (Romania); 
%MSHE (Russia); 
MICIU and AEI (Spain);
SNSF and SER (Switzerland); 
NASU (Ukraine); 
STFC (United Kingdom); 
DOE NP and NSF (USA).
%%%%%%%%%%%%%%%%%%%%%%%%%%%%%%%%%%%%%%%%%%%%%
We acknowledge the computing resources that are provided by ARDC (Australia), 
CBPF (Brazil),
CERN, 
IHEP and LZU (China),
IN2P3 (France), 
KIT and DESY (Germany), 
INFN (Italy), 
SURF (Netherlands),
Polish WLCG (Poland),
IFIN-HH (Romania), 
%RRCKI and Yandex LLC (Russia), 
PIC (Spain), CSCS (Switzerland), 
GridPP (United Kingdom),
and NSF (USA).  % added Feb2026
%%%%%%%%%%%%%%%%%%%%%%%%%%%%%%%%%%%%%%%%%% 
We are indebted to the communities behind the multiple open-source
software packages on which we depend.
%%%%%%%%%%%%%%%%%%%%%%%%%%%%%%%%%%%%%%%%%%
Individual groups or members have received support from
% ARC and ARDC (Australia); % moved to national 16/01/2025
RTP (Australia), % added 06/03/2026
Key Research Program of Frontier Sciences of CAS, CAS PIFI, CAS CCEPP (China); 
%Fundamental Research Funds for the Central Universities,  and Sci.\
%\& Tech.\ Program of Guangzhou (China); Removed 24/11/25
Minciencias (Colombia);
EPLANET, Marie Sk\l{}odowska-Curie Actions, ERC and NextGenerationEU (European Union);
A*MIDEX, ANR, IPhU and Labex P2IO, and R\'{e}gion Auvergne-Rh\^{o}ne-Alpes (France);
%RFBR, RSF and Yandex LLC (Russia);
Alexander-von-Humboldt Foundation (Germany);
ICSC (Italy); 
%GVA, XuntaGal, GENCAT, Inditex, InTalent and Prog.~Atracci\'on Talento, CM (Spain);
Severo Ochoa and Mar\'ia de Maeztu Units of Excellence, GVA, XuntaGal, GENCAT, InTalent-Inditex and Prog.~Atracci\'on Talento CM (Spain);
%XuntaGal --> Xunta de Galicia 
% SRC (Sweden);  % removed 27/02/2026 - end of grant
the Leverhulme Trust, the Royal Society and UKRI (United Kingdom).

\addcontentsline{toc}{section}{References}
%\setboolean{inbibliography}{true}
\bibliographystyle{LHCb/LHCb}
\bibliography{main,LHCb/standard,LHCb/LHCb-PAPER,LHCb/LHCb-CONF,LHCb/LHCb-DP,LHCb/LHCb-TDR,LHCb/LHCb-PUB}

\newpage
% \input{Authorship_LHCb-PAPER-2025-072}
% LHCb collaboration author list
% Data extracted on December 17th, 2025 at 8:59am for paper reference LHCb-PAPER-2025-072
\centerline
{\large\bf LHCb collaboration}
\begin
{flushleft}
\small
R.~Aaij$^{38}$\lhcborcid{0000-0003-0533-1952},
M. ~Abdelfatah$^{69}$,
A.S.W.~Abdelmotteleb$^{57}$\lhcborcid{0000-0001-7905-0542},
C.~Abellan~Beteta$^{51}$\lhcborcid{0009-0009-0869-6798},
F.~Abudin{\'e}n$^{59}$\lhcborcid{0000-0002-6737-3528},
T.~Ackernley$^{61}$\lhcborcid{0000-0002-5951-3498},
A. A. ~Adefisoye$^{69}$\lhcborcid{0000-0003-2448-1550},
B.~Adeva$^{47}$\lhcborcid{0000-0001-9756-3712},
M.~Adinolfi$^{55}$\lhcborcid{0000-0002-1326-1264},
P.~Adlarson$^{87}$\lhcborcid{0000-0001-6280-3851},
C.~Agapopoulou$^{14}$\lhcborcid{0000-0002-2368-0147}
% ...

{\footnotesize \it

% ...
$^{14}$Universit{\'e} Paris-Saclay, CNRS/IN2P3, IJCLab, Orsay, France\\
% ...
$^{38}$Nikhef National Institute for Subatomic Physics, Amsterdam, Netherlands\\
% ...
$^{47}$Instituto Galego de F{\'\i}sica de Altas Enerx{\'\i}as (IGFAE), Universidade de Santiago de Compostela, Santiago de Compostela, Spain\\
% ...
$^{51}$Physik-Institut, Universit{\"a}t Z{\"u}rich, Z{\"u}rich, Switzerland\\
% ...
$^{55}$H.H. Wills Physics Laboratory, University of Bristol, Bristol, United Kingdom\\
% ...
$^{57}$Department of Physics, University of Warwick, Coventry, United Kingdom\\
% ...
$^{59}$School of Physics and Astronomy, University of Edinburgh, Edinburgh, United Kingdom\\
$^{60}$School of Physics and Astronomy, University of Glasgow, Glasgow, United Kingdom\\
$^{61}$Oliver Lodge Laboratory, University of Liverpool, Liverpool, United Kingdom\\
% ...
$^{69}$Syracuse University, Syracuse, NY, United States\\
% ...
$^{87}$Department of Physics and Astronomy, Uppsala University, Uppsala, Sweden, associated to $^{60}$\\
% ...
\medskip
}
\end{flushleft}
% The author list for journal publications is generated from the
% Membership Database shortly after 'approval to go to paper' has been
% given.  It is available at \url{https://lbfence.cern.ch/membership/authorship}
% and will be sent to you by email shortly after a paper number
% has been assigned.  
% The author list should be included in the draft used for 
% first and second circulation, to allow new members of the collaboration to verify
% that they have been included correctly. Occasionally a misspelled
% name is corrected, or associated institutions become full members.
% Therefore an updated author list will be sent to you after the final
% EB review of the paper.  In case line numbering doesn't work well
% after including the authorlist, try moving the \verb!\bigskip! after
% the last author to a separate line.

% The authorship for Conference Reports should be ``The LHCb
% collaboration'', with a footnote giving the name(s) of the contact
% author(s), but without the full list of collaboration names.

% The authorship for Figure Reports should be ``The LHCb
% collaboration'', with no contact author and without the full list 
% of collaboration names.

\end{document}